\documentclass[10pt,aps,pre,reprint,twocolumn,amsmath,superscriptaddress,preprintnumbers,floatfix,nofootinbib,showpacs]{revtex4-1}
\usepackage{siunitx}
\usepackage{graphicx}
\usepackage[dvipsnames]{xcolor}
\usepackage[bookmarks]{hyperref} 
\usepackage{tikz}
\usepackage{booktabs}
\usepackage{mathtools}
\usepackage{braket}
\usepackage[utf8]{inputenc}
\usepackage[T1]{fontenc}
\usepackage{natbib}
\usepackage{amssymb,latexsym}
\usepackage{wasysym}
\usepackage{soul}
\usepackage{subfigure}
\usepackage{lipsum}

\DeclareMathOperator*{\argmin}{arg\,min}  
\begin{document}
\preprint{Draft for PRE \today}
\title{A new computational framework for particle and spin simulations \\ based on the stochastic Galerkin method}
\author{J.~Slim}
\affiliation{Institut f\"ur Hochfrequenztechnik, RWTH Aachen University, 52056 Aachen, Germany}
\author{F.~Rathmann}
\thanks{Corresponding author: f.rathmann@fz-juelich.de}
\affiliation{Institut f\"ur Kernphysik, Forschungszentrum J\"ulich, 52425 J\"ulich, Germany}
\author{D.~Heberling}
\affiliation{Institut f\"ur Hochfrequenztechnik, RWTH Aachen University, 52056 Aachen, Germany}
\affiliation{JARA--FAME (Forces and Matter Experiments), Forschungszentrum J\"ulich and RWTH Aachen University, Germany}

\begin{abstract}
An implementation of the Polynomial Chaos Expansion is introduced here as a fast solver of the equations of beam and spin motion inside an RF Wien filter. The device shall be used to search for the deuteron electric dipole moment in the COSY storage ring. The new approach is based on the stochastic Galerkin method, and it is shown that this constitutes a new and very powerful alternative to the commonly used Monte-Carlo methods. 
\end{abstract}

\maketitle

\section{Introduction and motivation} \label{int}
The JEDI\footnote{J\"ulich Electric Dipole moment Investigations, \\\url{http://collaborations.fz-juelich.de/ikp/jedi/}.} collaboration aims at a measurement of the electric dipole moment (EDM) of charged hadronic particles, such as deuterons and protons. In the near future, a first direct EDM measurement of the deuteron will be carried out at the COoler SYnchrotron COSY, a conventional magnetic storage ring. To this end, an RF Wien filter\,\cite{Slim2016116} has recently been installed, which will be operated at some harmonic of the spin precession frequency, whereby the sensitivity to the deuteron EDM is substantially enhanced. 

In order to eliminate the false EDM signals, it is of crucial importance to understand the sources of magnetic imperfections in the accelerator ring\,\cite{PhysRevAccelBeams.20.072801}. Particle and spin tracking simulation constitute powerful tools to do so. Conducting a realistic simulation for a machine that includes mechanical, electrical and electromagnetic uncertainties of all of its elements is indeed a challenging task. The well-known Monte-Carlo (MC) simulation as a tool to conduct such a study is computationally very expensive. The dimension (number of random parameters) in such simulations is particularly large. To further clarify the complexity of the problem, just consider that the current in a dipole magnet is not perfectly uniform, and this will alter the magnetic fields, which will affect the spin motion and orbit motion of the particles. It is therefore important to quantify such uncertainties beforehand for the intended EDM experiment. In addition, many other factors like the mechanical fabrication tolerance and the accuracy of the alignment contribute as well.       

The principle problem with the MC method is the necessity to repeat the simulation \num{10000} or even tens of millions of times. For limited-budget projects such as ours, this is therefore not an option. Recently, as a first step toward a full systematic analysis of the future EDM experiments at COSY, we conducted a study to quantify the electromagnetic performance of the above mentioned RF Wien filter under mechanical uncertainties\,\cite{Slim201752}. This investigation made use of the so-called Polynomial Chaos Expansion (PCE)\,\cite{askeyWiener}, as an efficient and yet accurate alternative to the MC method.

In this paper, we present a computational framework to further reduce the complexity of the problem in terms of the required number of tracking simulations. The same methodology as in Ref.\,\cite{Slim201752} is used here, but as we have access to the equations governing the system, here an \textit{intrusive} version of the PCE is employed.

This paper is organized as follows. Section\,\ref{sec:pce} briefly reviews the basic theoretical foundations of the PCE. Section\,\ref{sec:sgm} introduces the stochastic Galerkin method (SGM) and applies it to the beam and spin equations. Section\,\ref{sec:sim} describes the main steps required to perform the SGM. Section\,\ref{sec:res} presents the results  and compares them quantitatively to the equivalent MC results. Section\,\ref{sec:conc} summarizes and concludes the paper.    

\section{Polynomial chaos expansion} \label{sec:pce}
The polynomial chaos expansion (PCE) is a stochastic spectral method that allows for stochastically varying physical entities  $\mathcal{Y}$, as a response of some random input $\xi$  to be represented in terms of orthogonal polynomials. PCE permits $\mathcal{Y}$ to be expanded into a series of orthogonal polynomials of degree $p$ (the expansion order) as function of the input variables $\xi$. Thus it follows that
\begin{equation}
\mathcal{Y}=\sum_i^N a_i \Psi_i(\xi).
\label{eq:main_pce}
\end{equation}

Here the $\Psi_i$ are the multivariate orthogonal polynomials (the basis functions) of degree $p$,  and the $a_i$ are the expansion coefficients to be computed. The orthogonal polynomials can be the Hermite, Legendre, Laguerre, or any other  set of orthogonal polynomials, depending on the probabilistic distribution of the random input variables $\xi$. This paper deals with physical entities $\mathcal{Y}$ that include electromagnetic fields with uncertainties, particle positions, velocities and spin vectors. 

The first step in building a polynomial chaos series is to determine the probabilistic distribution of the input random variables and their number (the dimensionality of the problem). Once known, the multivariate orthogonal polynomials can be constructed using the three-terms recurrence properties that can be found \textit{e.g.}, in \,\cite{Parthasarathy20061083}. 

Oftentimes, the number of random variables is larger than unity, which implies that multivariate polynomials must be used instead of univariate ones. Obviously, the expansion order must be known and, as explained in\,\cite{blatman2010adaptive,yu2015advanced}, can be changed adaptively during processing. 

The expansion coefficients can be calculated using intrusive and non-intrusive methods. Non-intrusive methods consider the deterministic code as a black box, \textit{i.e.}, they do not alter the code nor the equations. The expansion coefficients are calculated using multiple calls to the deterministic code either via \textit{projection} or \textit{regression}. Both require a number of $N$ realization pairs $\left(\xi_i,\Psi_i \right)$ (see Eq.\,(\ref{eq:main_pce})). \textit{Projection} requires the evaluation of expectation values and relies on the orthogonality of polynomials to compute the coefficients in the form of
\begin{equation}
  a = \frac{ \mathbb{E} \left\{\mathcal{Y}\Psi \right\}}{\mathbb{E} \left\{\Psi^2 \right\}} \,,
  \label{eq:nisp}
\end{equation} 
where the computation of the expectation values ($\mathbb{E}\{\cdot\}$) necessitates the evaluation of integrals. Quadrature methods are one way to do so, and are commonly used in PCE analyses. Depending on the type of input distribution, the corresponding quadrature rule can be used. The Gauss-Laguerre quadrature for instance\,\cite{Abramowitz:1974:HMF:1098650}, is used in the case of uniformly distributed random variables. This projection method is widely known as non-intrusive spectral projection (NISP)\,\cite{LeMaitre:1339414}.

\textit{Regression}, on the other hand, estimates the coefficients that minimize the functional difference between the estimated response $\hat{\mathcal{Y}}$ and the actual response ${\mathcal{Y}}$, given by
\begin{equation}
a = \argmin \left( \mathbb{E}\left\{\hat{\mathcal{Y}}-\mathcal{Y}\right\}^2\right)\,.
\label{eq:reg_func}
\end{equation} 
The solution of Eq.\,(\ref{eq:reg_func}), obtained by linear regression, yields
\begin{equation}
a_i =\left(\Psi^T \cdot\Psi \right)\cdot\Psi\cdot \mathcal{Y}.
\label{eq:reg}
\end{equation}

In the context of this work, the regression method\,\cite{Offermann2015} provides more accurate results than the projection method, it is therefore used here to calculate the expansion coefficients $a_i$, which are subsequently fed into the stochastic Galerkin solver.

\section{Stochastic Galerkin method applied to beam and spin dynamics} \label{sec:sgm}
\subsection{Beam dynamics} \label{sec:beam-dynamics}
One of the most common methods to solve differential equations is the Galerkin-finite element method (FEM)\,\cite{zienkiewicz2013finite,Jin:2014:FEM:2655347}. In 1921 Boris Galerkin, a Russian mathematician, proposed a method to solve differential equations based on functional analysis. In contrast to other methods, such as finite difference (FD) schemes, the Galerkin method does not solve the differential equations directly, but transforms them into a variational form (a functional) that is then minimized. The functions minimizing this functional are the solutions to the required differential equations. The variational form is constructed via the Galerkin projection techniques [insert proper ref]. 

In this section, the construction of the variational form of the beam and spin dynamic equations using the stochastic Galerkin projection is described in detail. Neglecting forces other than the electromagnetic ones acting on the charged particles, the beam equations read
\begin{equation}
\begin{split}
\frac{d}{dt} \vec{v} & =\frac{q}{m\gamma} \left[ \vec{E} + \vec{v} \times \vec{B} -\frac{1}{c^2}\vec{v} \left(\vec{v} \cdot \vec{E} \right) \right]\,,\text{ and} \\
\frac{d}{dt} \vec{r} & =\vec{v} \,.
\end{split}
\label{eq:eom}
\end{equation}

\begin{figure*}[htb]
	\centering
	\subfigure[$C_{14jl}$]{\includegraphics[width=0.18\textwidth]{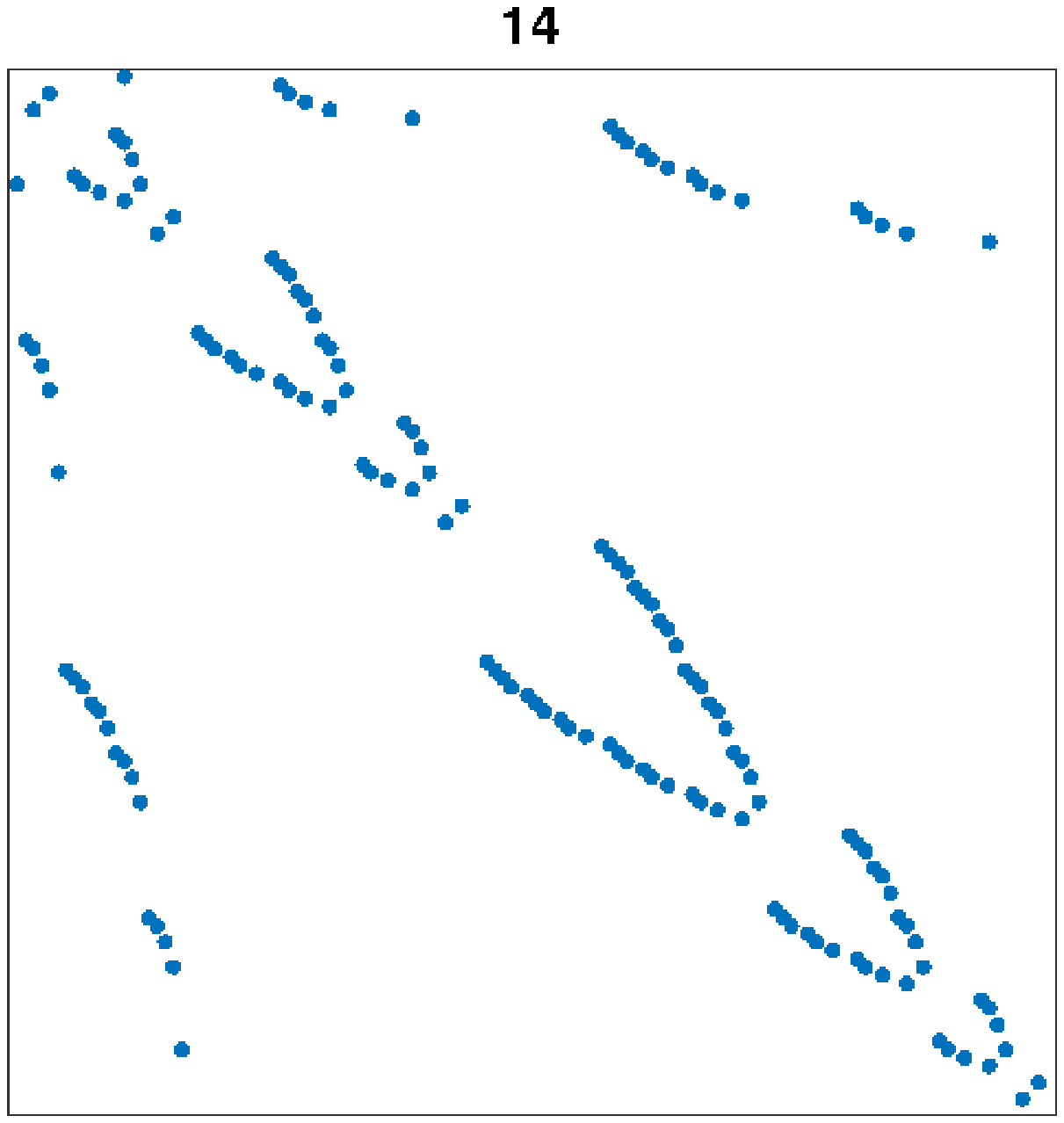}} \hspace{2mm}
	\subfigure[$C_{26jl}$]{\includegraphics[width=0.18\textwidth]{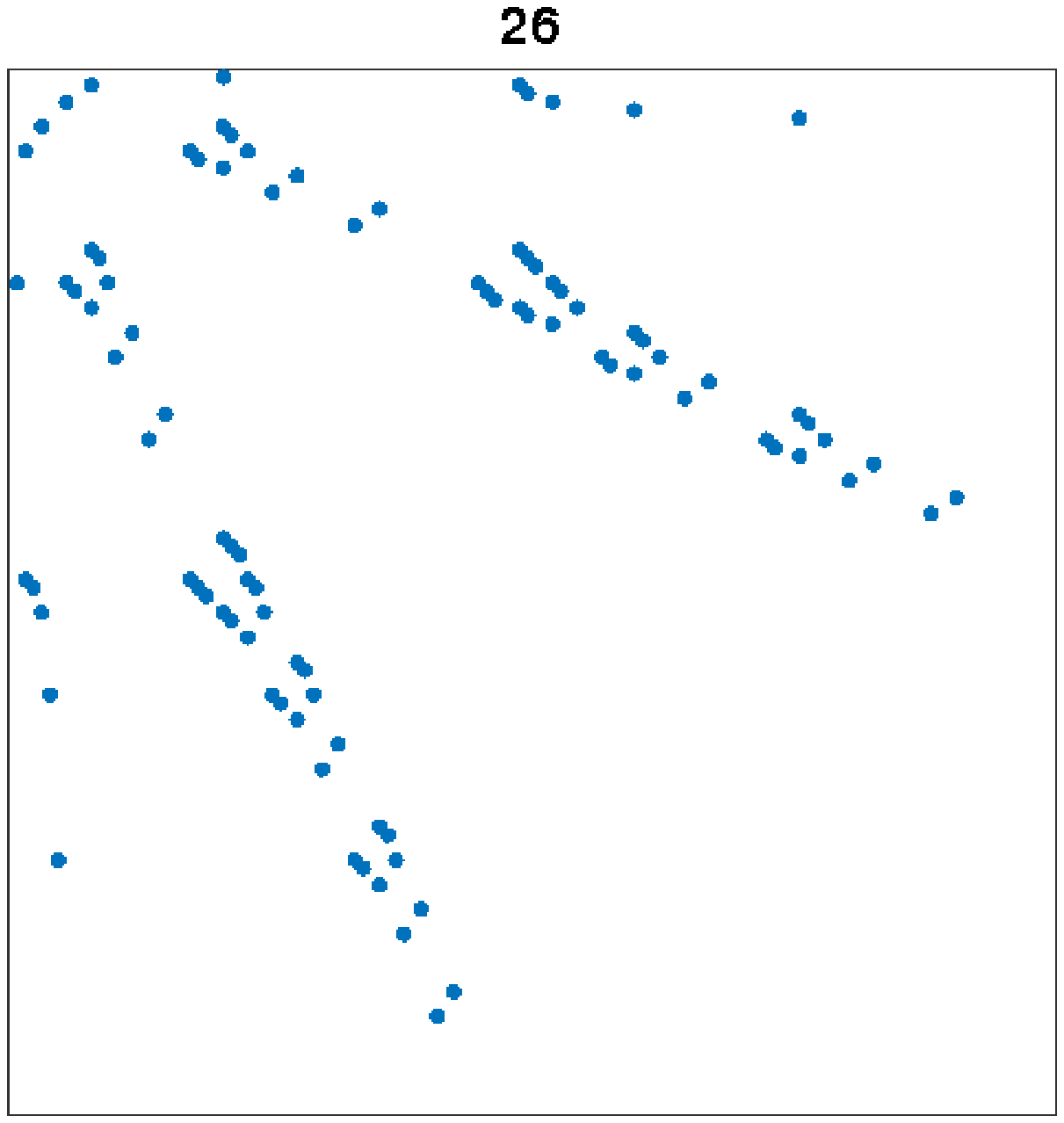}} \hspace{2mm}
	\subfigure[$C_{37jl}$]{\includegraphics[width=0.18\textwidth]{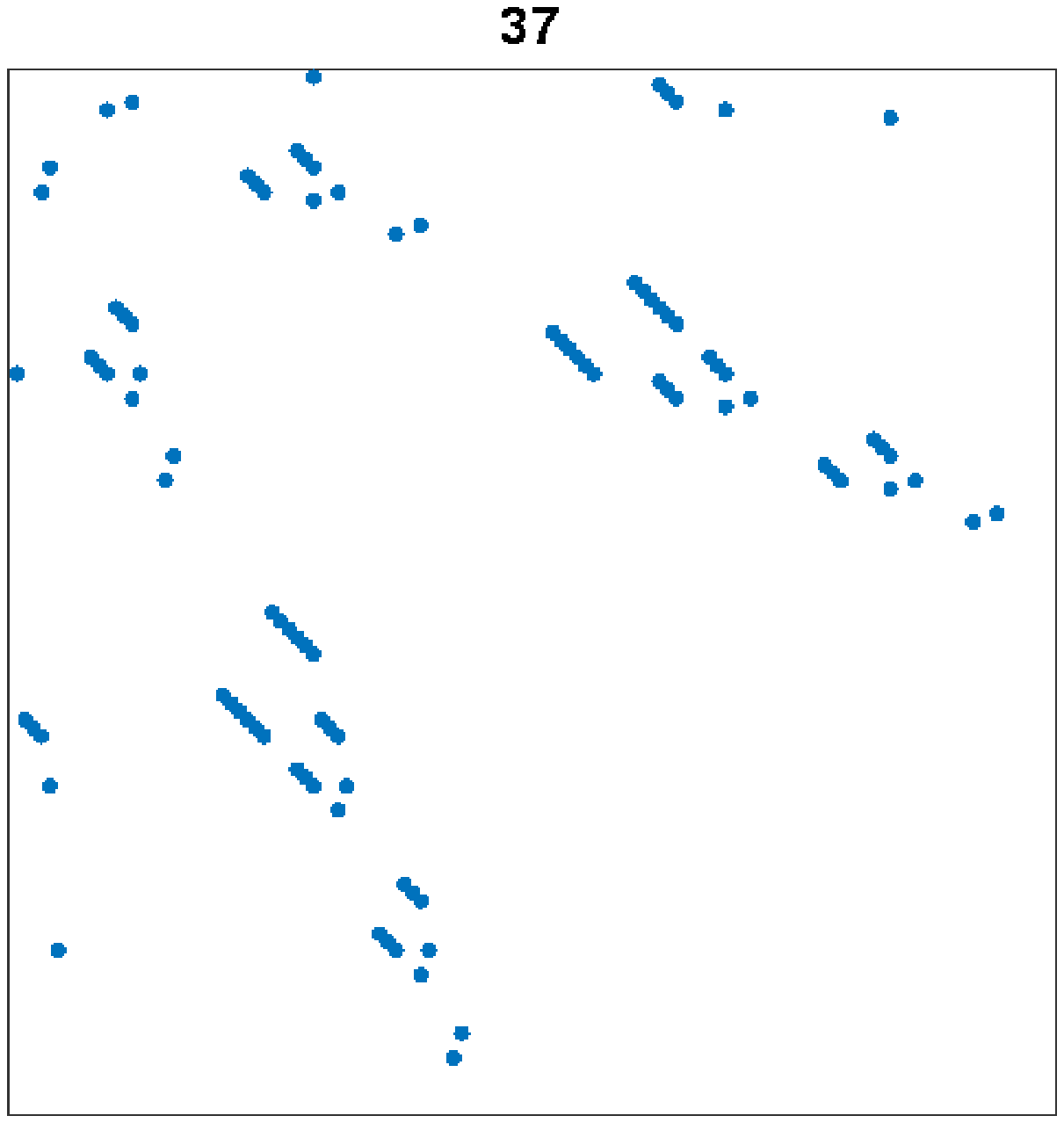}} \hspace{2mm}
	\subfigure[$C_{45jl}$]{\includegraphics[width=0.18\textwidth]{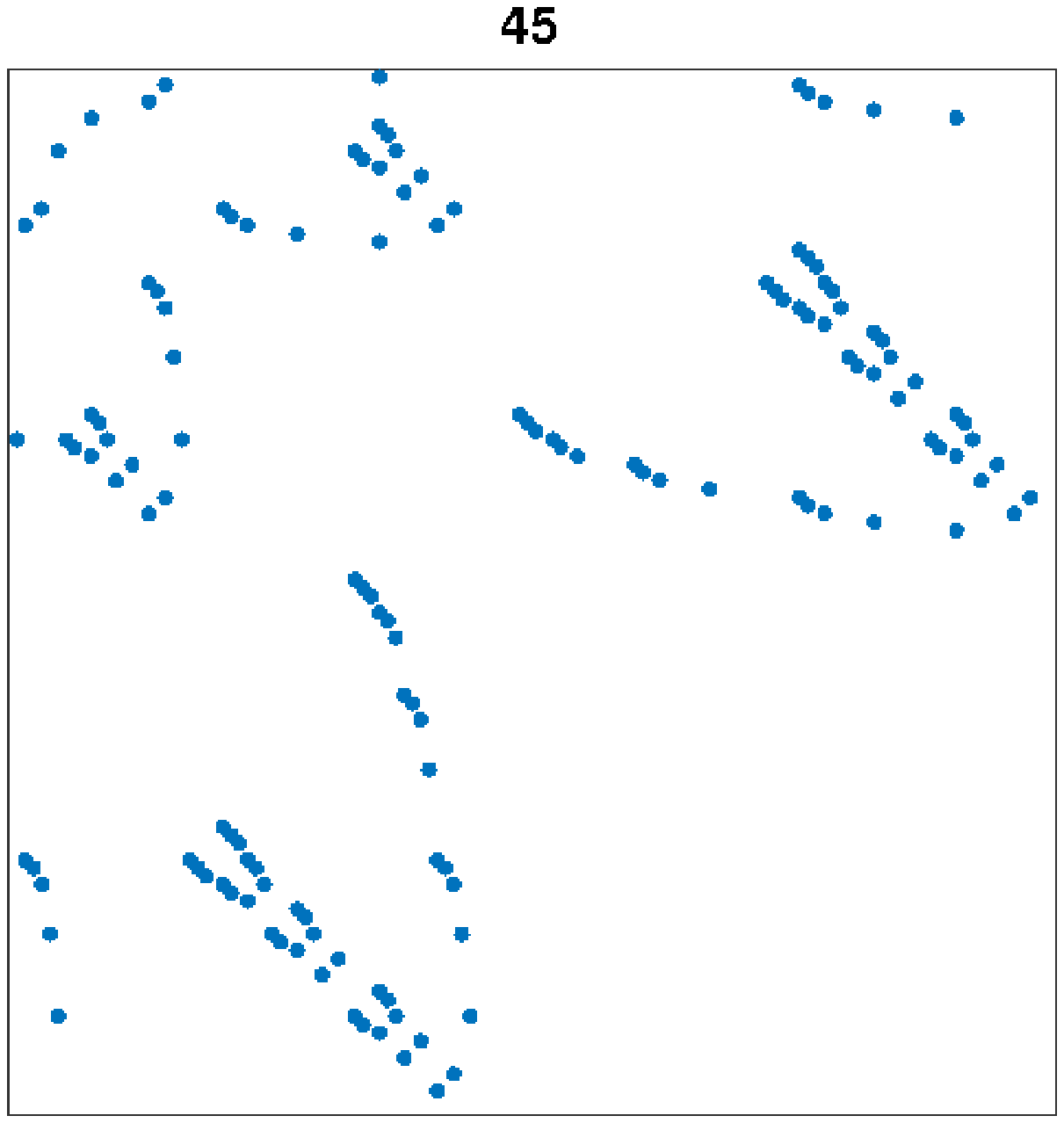}} \hspace{2mm}
	\subfigure[$C_{69jl}$]{\includegraphics[width=0.18\textwidth]{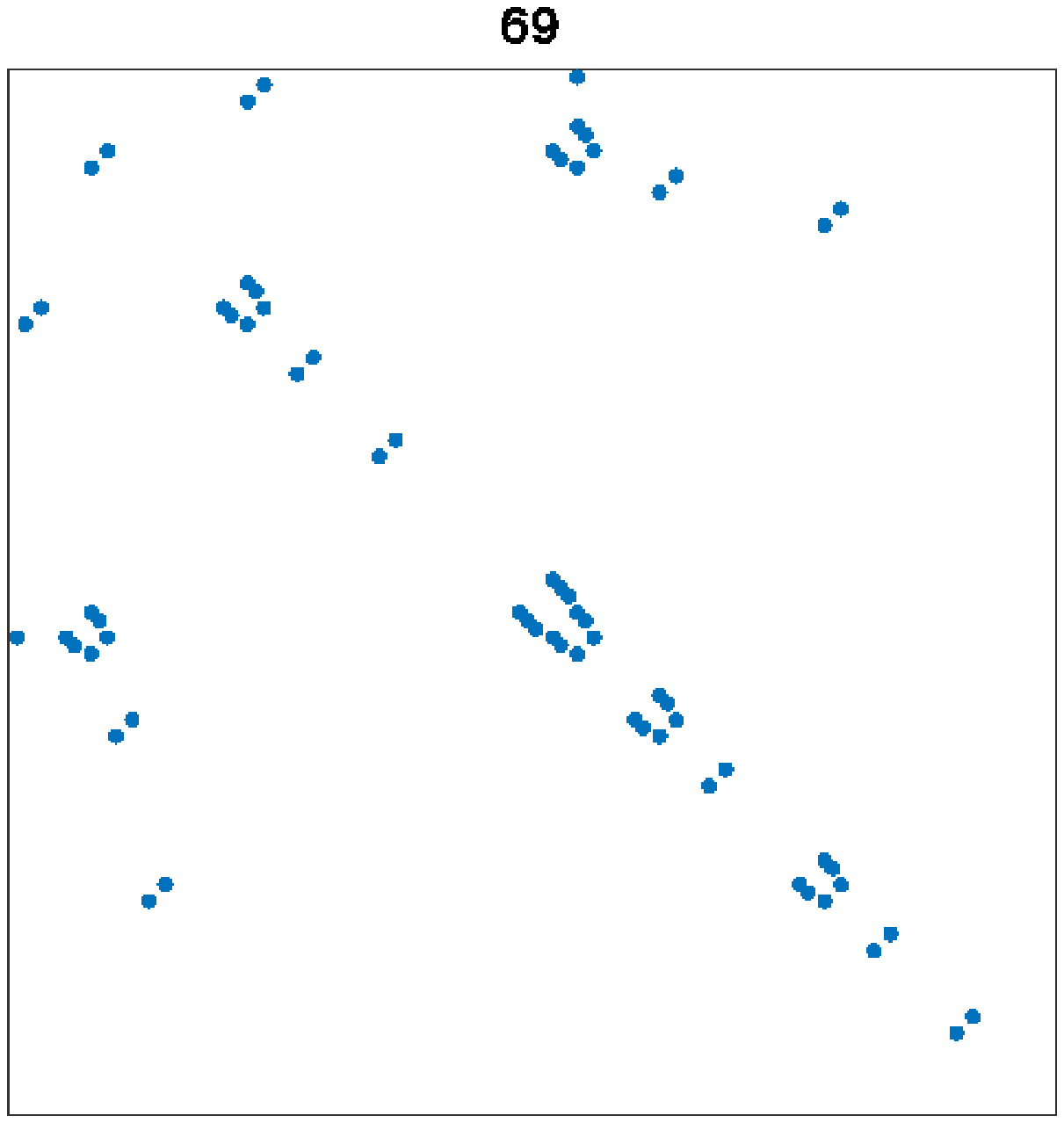}}\\
	\subfigure[$C_{77jl}$]{\includegraphics[width=0.18\textwidth]{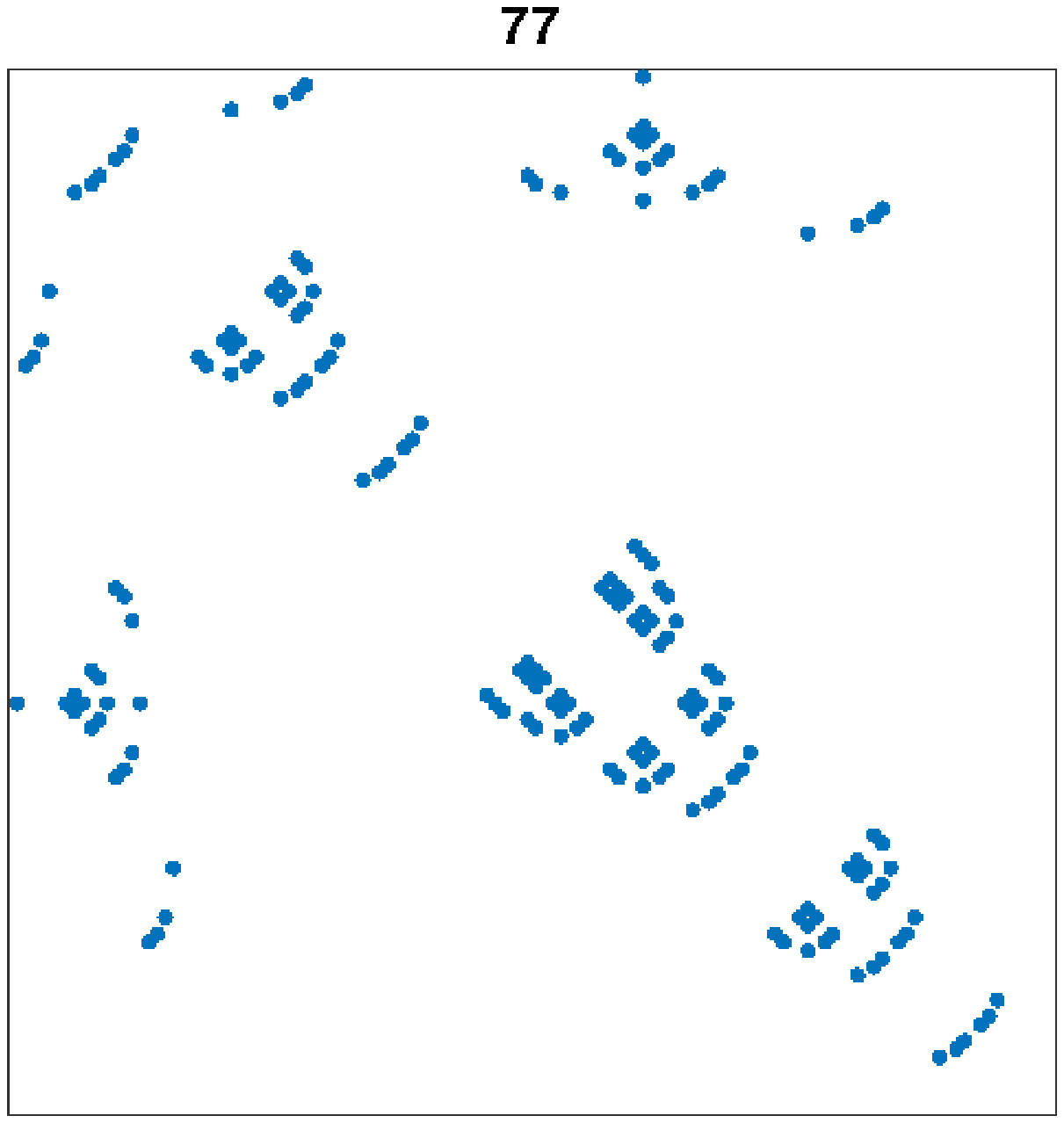}} \hspace{2mm}
	\subfigure[$C_{86jl}$]{\includegraphics[width=0.18\textwidth]{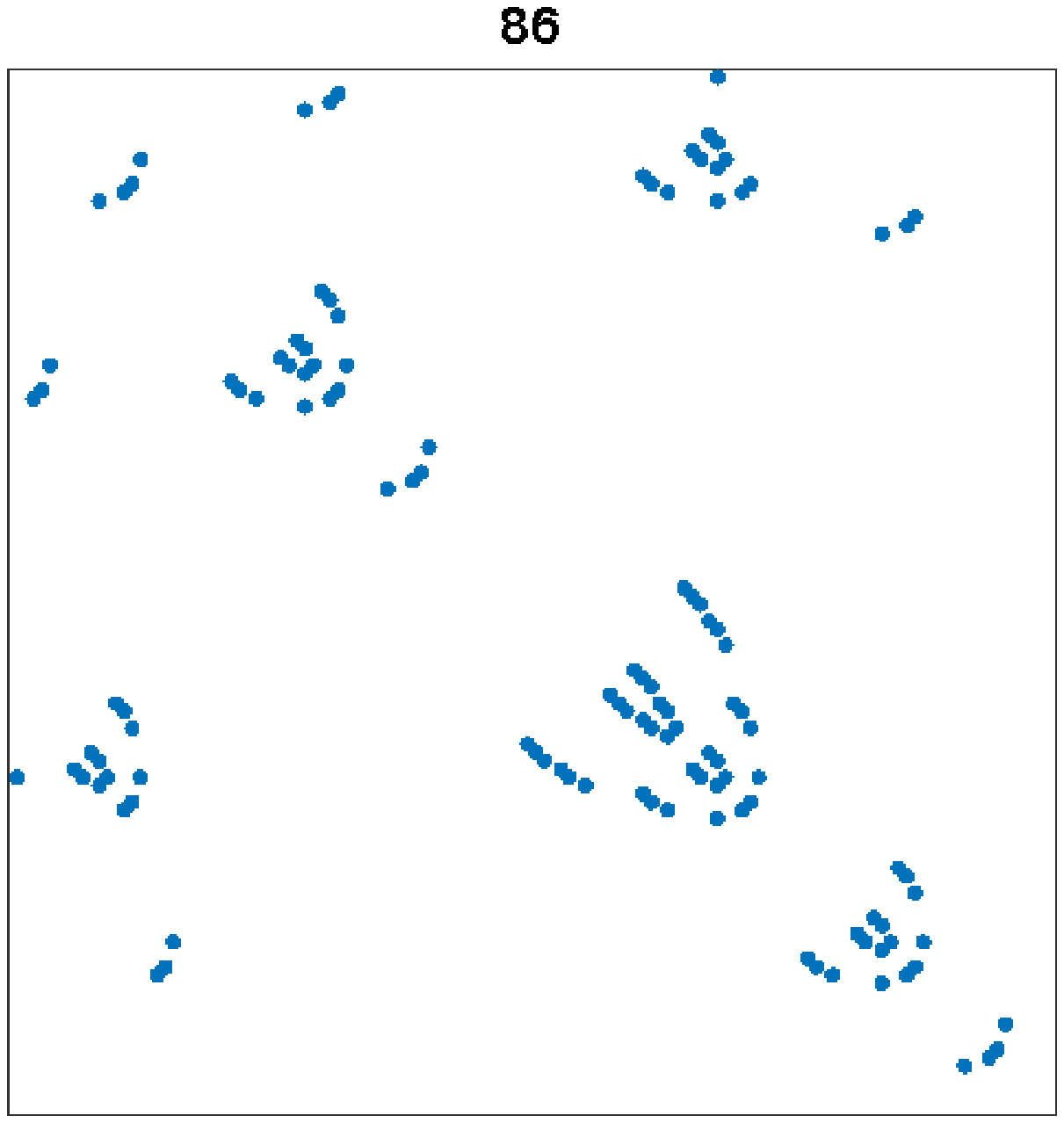}} \hspace{2mm}
	\subfigure[$C_{93jl}$]{\includegraphics[width=0.18\textwidth]{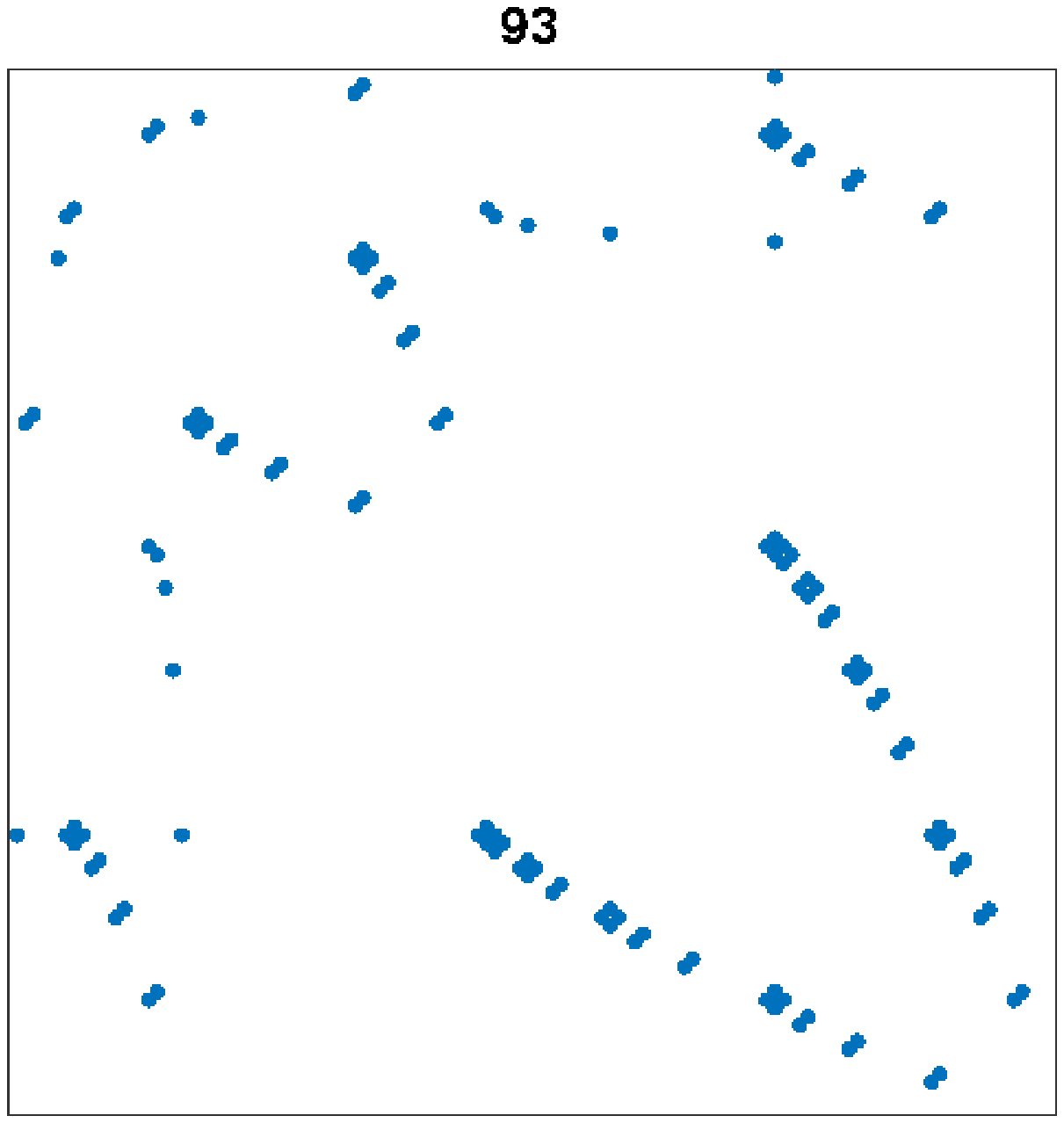}} \hspace{2mm}
	\subfigure[$C_{105jl}$]{\includegraphics[width=0.18\textwidth]{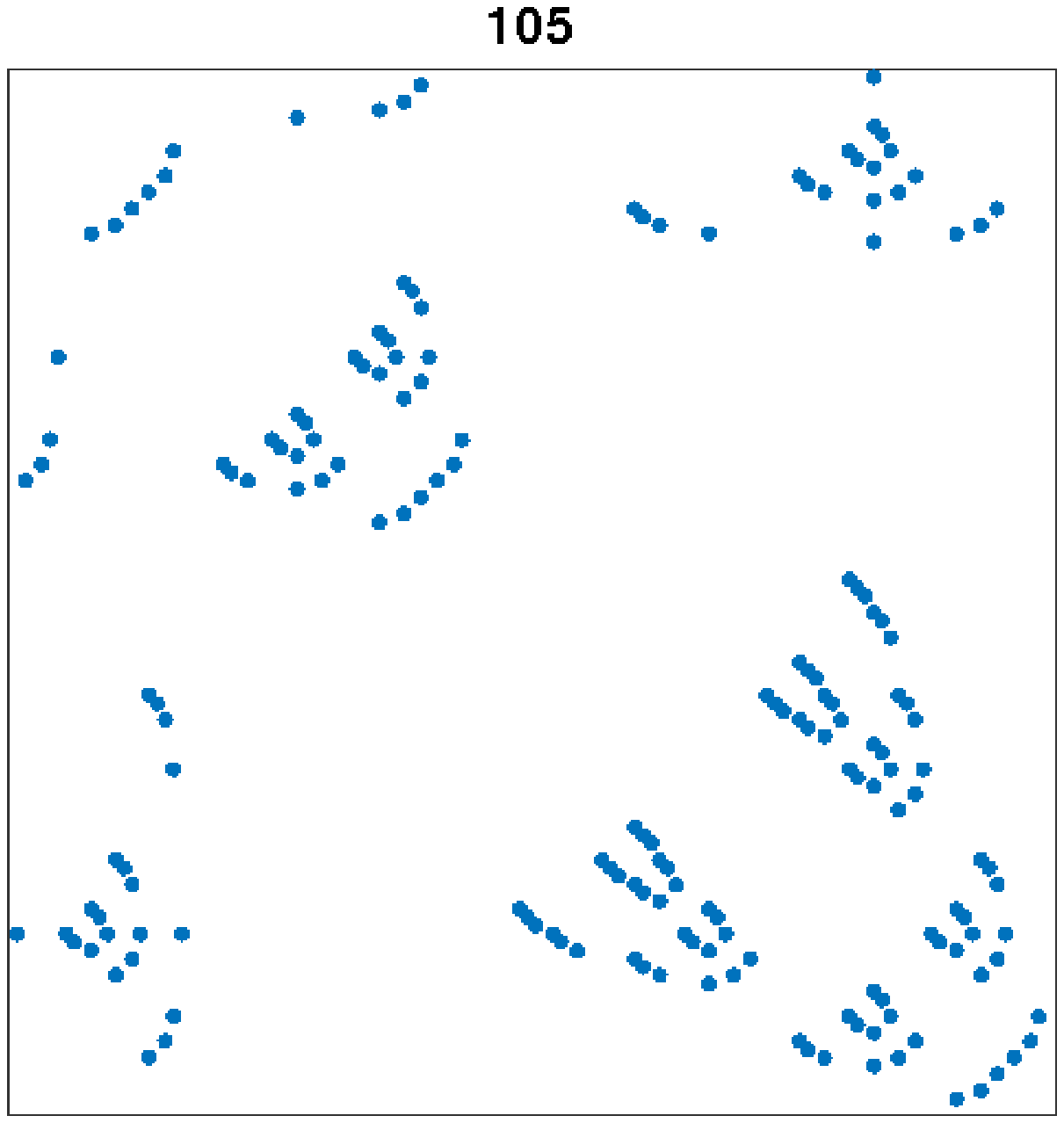}} \hspace{2mm}
	\subfigure[$C_{116jl}$]{\includegraphics[width=0.18\textwidth]{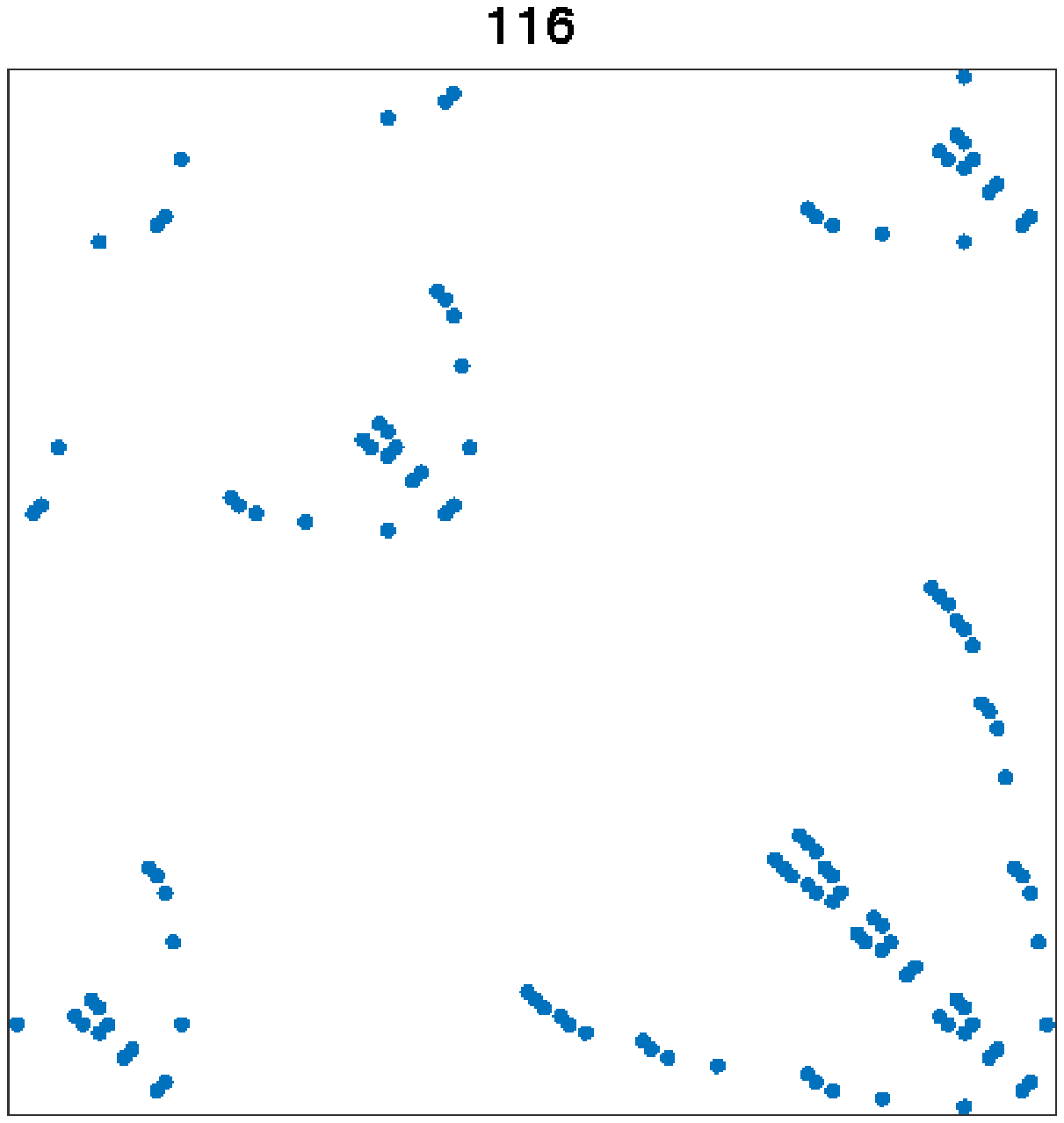}} 
	\caption{
	With the $m= 5$ dimensional problem and an expansion order of $p=4$, the number of basis functions is $P=126$ (see Eq.\,(7) of\,\cite{Slim201752}), which results in a $(126 \times 126 \times 126$) $C_{ijl}$ tensor. The sparsity of the rank-3 tensor $C_{ijl}$ is illustrated by fixing, \textit{e.g.},  the first index to values of $i=14$, $26$, $37$, $45$, $69$, $77$, $86$, $93$, $105$, and $116$, which yields the \num{10} sparse matrices shown. The multiplication of the PCE coefficients involving this tensor is very fast, as most of the tensor elements are zero, while computing, storing and loading of this tensor constitutes a CPU and memory intensive operation.
	\label{fig:c_ijl}}
\end{figure*}

The expansion of Eq.\,(\ref{eq:eom}) in Cartesian coordinates yields a linear system of six coupled ordinary differential equations,
\begin{subequations}
	\begin{align}
		 & \frac{d}{dt} v_x	=  \frac{q}{m} \left[ \frac{1}{\gamma} E_x+ \frac{1}{\gamma}v_y B_z - \frac{1}{\gamma}v_z B_y - \frac{1}{c^2 \gamma} v_x(\vec{v}\cdot \vec{E}) \right ] \,, \label{eq:vx}           \\
		 & \frac{d}{dt} v_y	=  \frac{q}{m} \left[ \frac{1}{\gamma}E_y+\frac{1}{\gamma} v_z B_x - \frac{1}{\gamma}v_x B_z - \frac{1}{c^2 \gamma}v_y(\vec{v}\cdot \vec{E})\right] \,, \label{eq:vy}               \\
		 & \frac{d}{dt} v_z	=  \frac{q}{m} \left[ \frac{1}{\gamma}E_z+\frac{1}{\gamma} v_x B_y - \frac{1}{\gamma}v_y B_x - \frac{1}{c^2 \gamma}v_z(\vec{v}\cdot \vec{E})\right] \,, \text{ and } 	\label{eq:vz} \\
		 & \frac{d}{dt} x 		=  v_x \,,								\label{eq:x}                                                                                                                                                                \\
		 & \frac{d}{dt} y 		=  v_y \,,	 							\label{eq:y}                                                                                                                                                               \\
		 & \frac{d}{dt} z 		=  v_z \,.								\label{eq:z}
	\end{align}
	\label{eq:eom_cart}
\end{subequations}
\noindent Here, $\vec{v}$ denotes the velocity vector of the particles, $q$  the particle charge, $m$ the mass, $\gamma$ the Lorentz factor, and $\vec{r}$ the position vector. 

When intra-beam scattering and other collective effects are neglected, the simulation of a beam of particles is equivalent to individual simulations with different initial conditions. It is furthermore assumed that the number of single particle simulations is sufficiently large to describe the beam. These assumptions permit us to use stochastic methods to solve the differential equations with random coefficients, or with uncertain input variables, or even with random boundary values. According to a probabilistic distribution, each individual particle of the population has a different initial position $\vec{r}$ and velocity vector $\vec{v}$. Therefore, the treatment of these parameters as random vectors in Eq.\,(\ref{eq:eom}) justifies the application of the SGM.

Equation\,(\ref{eq:eom}) describes an initial value problem, where the initial values vary randomly. The initial values are expanded using non-intrusive PC, particularly with linear regressions, and then the SGM \cite{xiu2010numerical} is applied to solve Eq.\,(\ref{eq:eom}). As an example, the technique of solving for the variable $v_x$ is discussed in detail below. 

$v_x$ is expanded as
\begin{equation}
v_x= \sum_i^N {v_{x}}_i^{(k)}\Psi_i \,,
\label{eq:vx_1}
\end{equation}  
where the ${v_{x}}_i^{(k)}$ are the chaos expansion coefficients. The superscript ${(k)}$ is used to identify the expansion coefficients, and also to emphasize that the variables are discretized. The coefficients are calculated according to 
\begin{equation}
{v_{x}}_i^{(k)} =\left( \Psi^T \cdot\Psi \right)\cdot\Psi\cdot {v_x}_{_0}\,,
\label{eq:reg_vx}
\end{equation}
where ${v_x}_{_0}$ are the initial \textit{x}-components of the particle velocities. Plugging Eq.\,(\ref{eq:vx_1}) into the left-hand side of Eq.\,(\ref{eq:vx}), we find
\begin{equation}
\frac{d}{dt} v_x =  \frac{d}{dt}\sum_i^N {v_{x}}_i^{(k)} \Psi_i =  \sum_i^N \frac{d}{dt} {v_{x}}_i^{(k)} \Psi_i \,.
\label{eq:vx_2}
\end{equation}

Now, the \textbf{stochastic Galerkin projection} is applied by multiplying Eq.\,(\ref{eq:vx_2}) with $\Psi_l$ and taking the expectation value $\mathbb{E}\{\cdot\}$, which gives
\begin{equation}
\begin{split}
	\mathbb{E} \left\{ \sum_i^N \frac{d}{dt} {v_{x}}_i^{(k)} \Psi_i \Psi_l \right\} & =\sum_i^N \frac{d}{dt} {v_{x}}_i^{(k)} \mathbb{E}\bigg\lbrace \Psi_i \Psi_l \bigg\rbrace \\
	                                                                                & =\sum_i^N \frac{d}{dt} {v_{x}}_i^{(k)} \langle \Psi_i \Psi_l \rangle                     \\
	                                                                                & =\sum_i^N\frac{d}{dt} {v_{x}}_i^{(k)} \langle \Psi_i^2\rangle \delta_{il} \,,
\end{split}
\end{equation}
where $\delta_{il}$ is the Kronecker delta which results from the orthogonality of the polynomials.

The electric field is also represented stochastically\footnote{The Cartesian components of the electric and magnetic fields ($\vec{E}$ and $\vec{B}$) are functions of the position vector $\vec{r}$, \textit{e.g.},  $\vec{E}_x \left(\vec{r}\,\right)$, $\vec{B}_x\left(\vec{r}\,\right)$, etc. The dependence of the field components, for instance  $\vec{E}_x \left(\vec{r}\,\right) =\vec{E}_x \left(x,y,z\right)$, on position, does not pose a problem for the PCE method, as long as the input variables (\textit{e.g.}, $\vec{r}$ and $\vec{v}$) are independent. The field components are considered as a black box and the expansion is carried out non-intrusively.} by a finite series as
\begin{equation}
E_x=  \sum_i^N {e_{x}}_i^{(k)} \Psi_i \,.
\label{eq:ex}
\end{equation}

The Lorentz factor $\gamma$ constitutes also a stochastic variable. Unfortunately, it appears in the denominator in all terms of Eq.\,(\ref{eq:eom}). To solve this problem, $1/\gamma$ is expanded instead of $\gamma$. Let $\alpha$ be defined as
\begin{equation}
\alpha = \frac{1}{\gamma}\,,
\end{equation}
then $\alpha$ is expanded as
\begin{equation}
\alpha=  \sum_i^N {\alpha}_i^{(k)} \Psi_i \,.
\label{eq:oogamma}
\end{equation}

The stochastic Galerkin projection is applied by multiplying the product of Eqs.\,(\ref{eq:ex}) and (\ref{eq:oogamma}) by $\Psi_k$, subsequently the expectation value $\mathbb{E}\{\cdot\}$ is calculated. It thus follows,
\begin{equation}
\begin{split}
	\mathbb{E} \left\{ \sum_i^N {e_{x}}_i^{(k)} {\Psi}_i \sum_j^N {\alpha}_j^{(k)} {\Psi}_j {\Psi}_l \right\} & =\sum_i^N \sum_j^N {e_{x}}_i^{(k)} {\alpha}_j^{(k)} \langle {\Psi}_i {\Psi}_j {\Psi}_l\rangle \\
	                                                                                                           & =\sum_i^N  \sum_j^N {\alpha}_i^{(k)} {e_{x}}_j^{(k)} C_{ijl}.
\end{split}
\end{equation}

The $C_{ijl} = \langle {\Psi}_i {\Psi}_j {\Psi}_l\rangle$ constitutes a sparse rank-3 tensor. It is constructed offline by computing the tensor product which constitutes a CPU intensive operation. The formula to compute $C_{ijl}$, provided in\,\cite{xiu2010numerical}, works only for low order and low-dimensional cases. In addition, it is limited to Gaussian-distributed random variables, and consequently applies only to Hermite polynomials. For the actual version implemented here, $C_{ijl}$ is computed numerically and distribution-independent. The implementation has been validated with a one-dimensional quadrature-based one and yielded the same results. 

Fortunately, $C_{ijl}$ needs to be computed only once. It can be stored and re-used when required. Although the multiplications of the PCE coefficients involve the $C_{ijl}$ term, this arithmetic operation does not introduce any computational overhead as $C_{ijl}$ is sparse. This is illustrated in Fig.\,\ref{fig:c_ijl} for several typical examples of the $C_{ijl}$ tensor .

The product of $\alpha$, magnetic field and velocity is more complicated, because it involves multiple polynomials. The triple product of $\alpha$, $v_y$ and $B_z$ requires to expand the two latter quantities as
\begin{equation}
\begin{split}
v_y &= \sum_i^N {v_y}_i^{(k)}\Psi_i \, ,\\
B_z &= \sum_i^N {b_{z}}_i^{(k)} \Psi_i \, ,
\end{split}
\end{equation}
\noindent
where ${v_y}_i^{(k)}$ and ${b_{z}}_i^{(k)}$ are the expansion coefficients of $v_y$ and $B_z$, respectively. The multiplication of the three sums yields
\begin{equation}
 \alpha v_y B_z = \sum_i^N \sum_j^N \sum_k^N  {\alpha}_i^{(k)}{v_{y}}_j^{(k)}{b_{z}}_k^{(k)} \Psi_i \Psi_j \Psi_k\,.
\label{eq:vybz}\end{equation}

By applying the stochastic Galerkin projection to Eq.\,(\ref{eq:vybz}), it follows that
\begin{equation}
\begin{split}
	\mathbb{E} \left\{ \alpha v_y B_z \Psi_l \right\} & = \sum_i^N \sum_j^N \sum_k^N  {\alpha}_i^{(k)} {v_{y}}_j^{(k)} {b_{z}}_k^{(k)} \langle\Psi_i \Psi_j \Psi_k \Psi_l \rangle \\
	                                                   & = \sum_i^N \sum_j^N \sum_k^N   {\alpha}_i^{(k)} {v_{y}}_i^{(k)} {b_{z}}_j^{(k)} D_{ijkl} \,.
\end{split}
\label{eq:rank-4-tensor}
\end{equation}
\noindent
$D_{ijkl}$ is similar to $C_{ijl}$, but it constitutes a rank-4 tensor. 

The case for the third term of Eq.\,(\ref{eq:vx}) yields
\begin{equation}
\begin{split}
	\mathbb{E} \left\{\alpha v_z B_y \Psi_k \right\} & =  \sum_i^N \sum_j^N \sum_k^N   {\alpha}_i^{(k)} {v_{z}}_j^{(k)} {b_{y}}_k^{(k)} \langle\Psi_i \Psi_j \Psi_k \Psi_l\rangle \\
	                                                  & = \sum_i^N \sum_j^N \sum_k^N   {\alpha}_i^{(k)} {v_{z}}_j^{(k)} {b_{y}}_k^{(k)} D_{ijkl} \,.
\end{split}
\end{equation}

The last term of the right hand side of Eq.\,(\ref{eq:vx}), \textit{i.e.},
\begin{equation}
\frac{1}{c^2} \alpha{v_x} \left( \vec{v} \cdot \vec{E} \right)
\label{eq:full-product}
\end{equation}
is yet more complicated, as it involves the scalar product. The scalar product operator multiplies the operands component-wise before summing them up. These operands, however, are PC coefficients. The corresponding multiplication is in fact a Galerkin one\footnote{We actually tried a simple multiplication, but the SGM did not converge to the correct solution, \textit{i.e.}, a strong variations of $x$, $y$, $v_x$, $v_y$ and $S_y$ with respect to the MC solution were observed.}, which involves a series of double products\,\cite{sullivan2015introduction}, given by
\begin{equation}
\vec{v}\cdot \vec{E} = \sum_i^3 \sum_j^N \sum_k^N {v_{i}}_j^{(k)} {e_{i}}_k^{(k)} \Psi_j \Psi_k \,.
 \label{eq:scalar}
\end{equation}

This means that Eq.\,(\ref{eq:scalar}) requires the stochastic Galerkin projection to compute a rank-5 tensor, which makes the Galerkin method highly inefficient\,\cite{sullivan2015introduction}. In order to solve this problem, a \textbf{pseudo-spectral method}\,\cite{LeMaitre:1339414,Pettersson2014481} is used. Using this method, the Galerkin projection is applied first to the auxiliary variable $g_k$ (the one representing the scalar product), and then secondly to the full product in Eq.\,(\ref{eq:full-product}). This way, the rank-4 tensor product, introduced above in Eq.\,\ref{eq:rank-4-tensor}), can be used. In particular, $g_k$ reads
\begin{equation}
g_{l}=\mathbb{E} \left\{ \left( \vec{v}\cdot \vec{E}\right)  \Psi_l \right\} =   \sum_i^3 \sum_j^N \sum_k^N {v_{i}}_j^{(k)} {e_{i}}_k^{(k)} C_{ijl} \,,
\end{equation}
where the subscript $l$ here constitutes a free variable. And then by applying the stochastic Galerkin projection it follows that 
\begin{equation}
\begin{split}
	\mathbb{E} \left\{ \alpha{v_x}\left(\vec{v}\cdot\vec{E}\right) \Psi_l \right\} = \sum_i^N \sum_j^N \sum_k^N   {\alpha}_i^{(k)} {v_{x}}_j^{(k)} {g}_k^{(k)} D_{ijkl} \,.
\end{split}
\label{eq:expectation-value-scalar-product}
\end{equation} 

Now that all the terms of Eq.\,(\ref{eq:vx}) are expanded, as described above, and after a similar treatment also Eqs.\,(\ref{eq:vy}) to\,(\ref{eq:z}), the system of ordinary differential equations (ODEs) has been solved.

\subsection{Spin dynamics}
The spin dynamics in an electromagnetic storage ring with non-vanishing EDM is described by the generalized T-BMT equation\,\cite{PhysRevLett.2.435,fukuyama}, which reads
\begin{equation}
\label{eq:spin}
\frac{d}{dt} \vec{S} = \left(\vec{\Omega}^{\text{MDM}} + \vec{\Omega}^{\text{EDM}}\right) \times \vec{S} \,.
\end{equation}
Here, $\vec{S}$ denotes the particles' spin, and $\vec{\Omega}^{\text{EDM}}$ and $\vec{\Omega}^{\text{MDM}}$ are the angular velocities associated with the magnetic (MDM) and the electric dipole moments (EDM).  $\vec{\Omega}^{\text{MDM}}$ and $\vec{\Omega}^{\text{EDM}}$ are defined as
\begin{equation}
\begin{split}
	\vec{\Omega}^{\text{MDM}}   = & -\frac{q}{{m} \gamma}  \bigg[ \left(1 + G \gamma\right)\vec{B} +  \left( G \gamma + \frac{\gamma}{1+\gamma}\right) \frac{\vec{E}\times \vec{\beta}}{c}                                           \\
	                              & - \frac{G \gamma^2}{\gamma+1} \vec{\beta}\left( \vec{\beta}\cdot \vec{B}\right)\bigg] \,,                                                                                                        \\
	\vec{\Omega}^{\text{EDM}}  =  & -\frac{q}{{m}}\frac{\eta}{2} \bigg[ \frac{\vec{E}}{c} + \vec{\beta}\times\vec{B}-\frac{\gamma}{\gamma+1}\vec{\beta}\left(\vec{\beta}\cdot\frac{\vec{E}}{c}\right)\bigg] \,.
\end{split}
\label{eq:omegas}
\end{equation}

The particle velocity is given by $\vec \beta=\vec{v}/ c$, $G$ denotes the anomalous magnetic moment, and $\eta$ is a dimensionless parameter that describes the particle EDM.

Expanding Eq.\,(\ref{eq:spin}) in Cartesian coordinates reads
\begin{equation}
\begin{split}
	\frac{d}{dt} S_x  = & \Omega_y^\text{MDM} S_z - \Omega_z^\text{MDM} S_y + \Omega_y^\text{EDM} S_z - \Omega_z^\text{EDM} S_y \,, \\
	\frac{d}{dt} S_y  = & \Omega_z^\text{MDM} S_x - \Omega_x^\text{MDM} S_z + \Omega_z^\text{EDM} S_x - \Omega_x^\text{EDM} S_z \,, \\
	\frac{d}{dt} S_z  = & \Omega_x^\text{MDM} S_y - \Omega_y^\text{MDM} S_x + \Omega_x^\text{EDM} S_y - \Omega_y^\text{EDM} S_x \,.
\end{split}
\label{eq:spin_cart}
\end{equation}

Before proceeding to the stochastic discretization of the spin equation in Eq.\,(\ref{eq:spin}), some variables are introduced to simplify the discretization process. Let
\begin{subequations}
	\begin{align}
		 & f_1	=  \frac{1}{\gamma} +G \,, 							 \label{eq:f1}                 \\
		 & f_2	=  \frac{1}{c} \left(G+\frac{1}{\gamma}\right) \,, \label{eq:f2} \\
		 & f_3	=  \frac{G \gamma}{1+\gamma} \,, \text{ and}					 \label{eq:f3}              \\
		 & f_4 	=  \frac{\gamma}{\gamma +1 } \,,					 \label{eq:f4}
	\end{align}
	\label{eq:spin_aux}
\end{subequations}

All the terms that interact with field and velocity components are grouped together. It should be noted that here only $\gamma$ constitutes a stochastic variable. The PCE coefficient are also calculated using the non-intrusive projection method, yielding
\begin{subequations}
	\begin{align}
		 & f_1	=  \sum_i^N {f_1}_i \Psi_i \,, 							 \label{eq:f1_pce}     \\
		 & f_2	=  \sum_i^N {f_2}_i \Psi_i \,,							 \label{eq:f2_pce}      \\
		 & f_3	=  \sum_i^N {f_3}_i \Psi_i \,,		\text{ and}					 \label{eq:f3_pce}      \\
		 & f_4 	=  \sum_i^N {f_4}_i \Psi_i \,.				    		  \label{eq:f4_pce}
	\end{align}	
	\label{eq:spin_aux_pce}
\end{subequations}

$\vec{\Omega}^{\text{MDM}}$ and $\vec{\Omega}^{\text{EDM}}$ are then rewritten as
\begin{equation}
\begin{split}
	\vec{\Omega}^{\text{MDM}}   = & -\frac{q}{{m}}  \bigg[ f_1\, \vec{B} +  f_2 \left( \vec{E}\times \vec{\beta}\right) - f_3\, \vec{\beta}\left( \vec{\beta}\cdot \vec{B}\right)\bigg] \,, \text{ and}                         \\
	\vec{\Omega}^{\text{EDM}}  =  & -\frac{q}{{m}}\frac{\eta}{2} \bigg[ \frac{\vec{E}}{c} + \vec{\beta}\times\vec{B}- f_4\, \vec{\beta}\left(\vec{\beta}\cdot\frac{\vec{E}}{c}\right)\bigg] \,.
\end{split}
\label{eq:omegas_f}
\end{equation}

In the following, the same methodology as described in Sec.\ref{sec:beam-dynamics}, is applied to the spin equation. Starting with the $x$-component of the spin vector in Eq.\,(\ref{eq:omegas}) by induction from the derivation described above, it follows that
\begin{equation}
\begin{split}
	\sum_i^N\frac{d}{dt} {S_x}^{(k)} = & \sum_i^N\sum_j^N {\Omega^{\text{MDM}}_y}_i^{(k)} {S_{z}}_j^{(k)} C_{ijl}    \\
	                                   & - \sum_i^N\sum_j^N {\Omega^{\text{MDM}}_z}_i^{(k)} {S_{y}}_j^{(k)} C_{ijl}  \\
	                                   & + \sum_i^N\sum_j^N {\Omega^{\text{EDM}}_y}_i^{(k)} {S_{z}}_j^{(k)} C_{ijl}  \\
	                                   & - \sum_i^N\sum_j^N {\Omega^{\text{EDM}}_z}_j^{(k)} {S_{y}}_j^{(k)} C_{ijl}\,.
\end{split}
\label{eq:spin2}
\end{equation}

Rewriting  ${\Omega^{\text{MDM}}_y}_i^{(k)}$ in terms of the individual components involved, for instance, is equivalent to the following expression
\begin{equation}
 \begin{split}
 	  & \sum_i^N {\Omega^{\text{MDM}}_y}_i^{(k)}                                                                                                                     \\
 	= & -\frac{q}{m} \bigg[ \sum_i^N  \sum_j {f_{1}}_i^{(k)} {b_{y}}_j^{(k)} C_{ijl}                                                                                 \\
 	  & + \sum_i^N  \sum_j \sum_k \left( {f_{2}}_i^{(k)} {e_z}_i^{(k)} {\beta_x}_j^{(k)} D_{ijkl} - {f_{2}}_i^{(k)} {e_x}_i^{(k)} {\beta_z}_j^{(k)} D_{ijkl} \right) \\
 	  & - \sum_i^N \sum_j^N \sum_k^N {f_{3}}_i^{(k)} {\beta_y}_j^{(k)} {h}_k^{(k)} D_{ijkl} \bigg] \,,
 \end{split}\label{eq:omega_y}
\end{equation}
where
\begin{equation}
h_{l}=\mathbb{E} \left\{ \left(\vec{\beta}\cdot \vec{B}\right)  \Psi_l \right\}=   \sum_i^3 \sum_j^N \sum_k^N {\beta_{i}}_j^{(k)} {b_{i}}_k^{(k)} C_{ijl} 
\end{equation}
Here, $l$ constitutes a dummy subscript [which is replaced by $k$ in Eq.(\ref{eq:omega_y})]. Similarly, the other components of the equation governing the spin dynamics [Eq.\,(\ref{eq:spin})] can be constructed, and this derivation is omitted here for brevity.

\section{Simulation walkthrough} \label{sec:sim}
The stochastic Galerkin method (SGM) transforms a system of differential equations describing the quantities of interest into an augmented system of equations containing only the coefficients. Depending on the dimension and the expansion order $p$, the dimension of the new system of equations is determined. The solutions of the system are called stochastic modes which are nothing else than the time- and position-dependent expansion coefficients. These coefficients can later on be used to reconstruct the response for an arbitrary number of particles. 

To perform the simulations, the random variables have to be identified first. This can include the particle positions, velocities, spins and the electric and magnetic fields. Next, the expansion order $p$ is selected with the smallest possible value in order for the augmented system of equations to be as small as possible as well. The expansion order can later be increased, in case the achieved accuracy is not satisfactory. In this work, the expansion order was set to ($p = 4$), which results in very good error values, as will be shown later. 

Since here, Gaussian distributions were selected, the basis functions are Hermite polynomials. The random variables must be standardized in order for the method to converge. The total number of polynomials (the cardinality of the polynomials) is $P = 126$\,\cite{Slim201752}. 

In particle simulations, particles are generated according to a well-defined phase-space population. An example is shown in Fig.\,9 of\,\cite{Slim2016116}, and such populations might violate the independence requirement of the random parameters. In this case, the Nataf transformation\,\cite{hurtado2013structural,kotulski2009error,LEBRUN2009172}\footnote{Nataf transformations are iso-probabilistic transformations that transform correlated Gaussian variables with arbitrary mean and variance into normally-distributed ones. Without such a transformation, the orthogonality of the basis functions would be violated.} can be used to solve this problem. 

With all these parameters, the expansion coefficients of the initial particle population are constructed using the linear regression method, as described in Eq.\,(\ref{eq:reg_vx}).

With $P = 126$, $C_{ijl}$ is a $126 \times 26 \times 126$ tensor. Unless a new random quantity is added to the analysis, or the order of the expansion is changed, the stored $C_{ijl}$ can be used when required. When the expansion order $p$ or the dimension of the problem is large, other methods, such as pseudo-spectral ones, may become more favorable (see \textit{e.g.}, \cite{Cottrill2011}). 

The SGM-based system of equations is solved using Matlab\footnote{Mathworks, Inc.\ Natick, Massachusetts, United States \url{http://www.mathworks.com}.}, employing the deterministic ordinary differential equations solver '\textit{ode45}' with a fixed time step of $\SI{1}{ms}$, and relative and absolute error tolerances of $10^{-13}$ and $10^{-20}$, respectively. The '\textit{ode45}' solver is based on the 4$^\text{th}$-order Runge-Kutta integration technique. It is important to note that the same ODE solver is used in both cases, MC and SGM. 

The implementation of the large system of equations was vectorized, and took less than a second (on average $\SI{0.47}{s} \pm \SI{0.2}{s}$) of CPU time\footnote{The simulations were performed on an HP Z840 workstation with a single Xeon E5v4 CPU and a RAM capacity of 80 GB.}. 

At the final stage, the performance of the SGM must be evaluated quantitatively, with the help of an adequate error analysis. Due to time and position dependence, the error calculation involves either the mean value or the standard deviation of the quantity under investigation, denoted by $\zeta$. The corresponding errors are called $\mathcal{\epsilon}_\mu$ and $\mathcal{\epsilon}_\sigma$, respectively, and are defined as
\begin{subequations}
\begin{align}
	\mathcal{\epsilon}_\mu(t)    & =	\bigg\lvert  \frac{\bar{\zeta}(t)-\bar{\hat{\zeta}}(t)}{\bar{\zeta}(t)}\bigg\rvert,\,\, \label{eq:err_mu}\text{and} \\
	\mathcal{\epsilon}_\sigma(t) & =	\bigg\lvert  \frac{  \sigma[\zeta(t)]-\sigma[\hat{\zeta}(t)]}{\sigma[\zeta(t)]}\bigg\rvert \label{eq:err_sigma}\,.
\end{align} \label{eq:err_time}
\end{subequations}

Here, $\zeta$ may refer to either the position, velocity or spin vector. $\hat{\zeta}$ denotes the estimated value using the SGM. To conduct an analysis similar to the one described in\,\cite{askeyWiener,Gerritsma20108333}, the exact initial conditions are fed to both the MC and the SGM solver, so that the solution can be compared on a particle-by-particle basis. In this way, the difference between the two solutions reflects the true performance of the  proposed new method.

\begin{figure*}[!]
	\centering
	\subfigure[Solution for $x(t)$.]   {\includegraphics[width=0.45\textwidth]{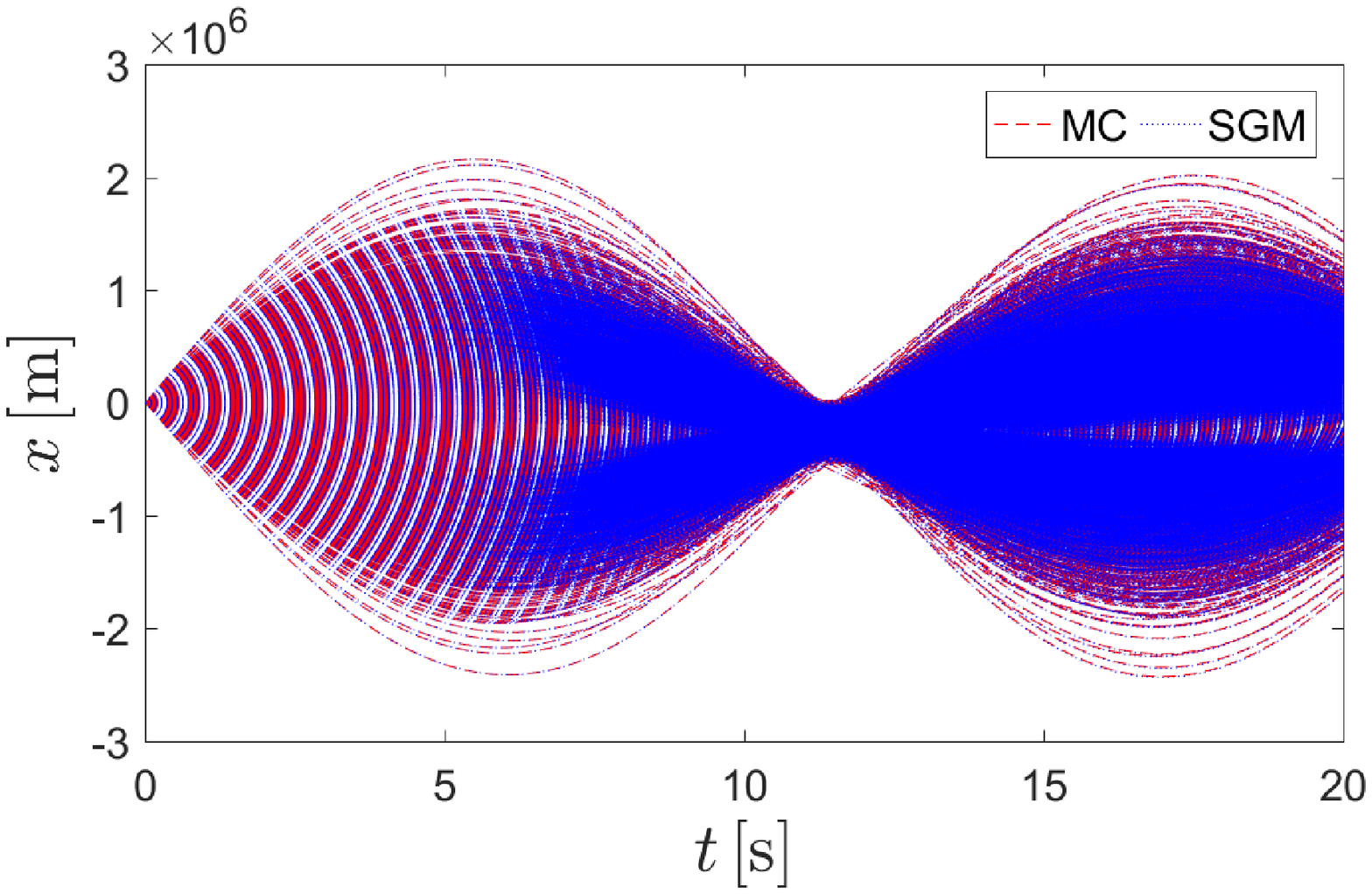}}
	\subfigure[Solution for $y(t)$.]   {\includegraphics[width=0.45\textwidth]{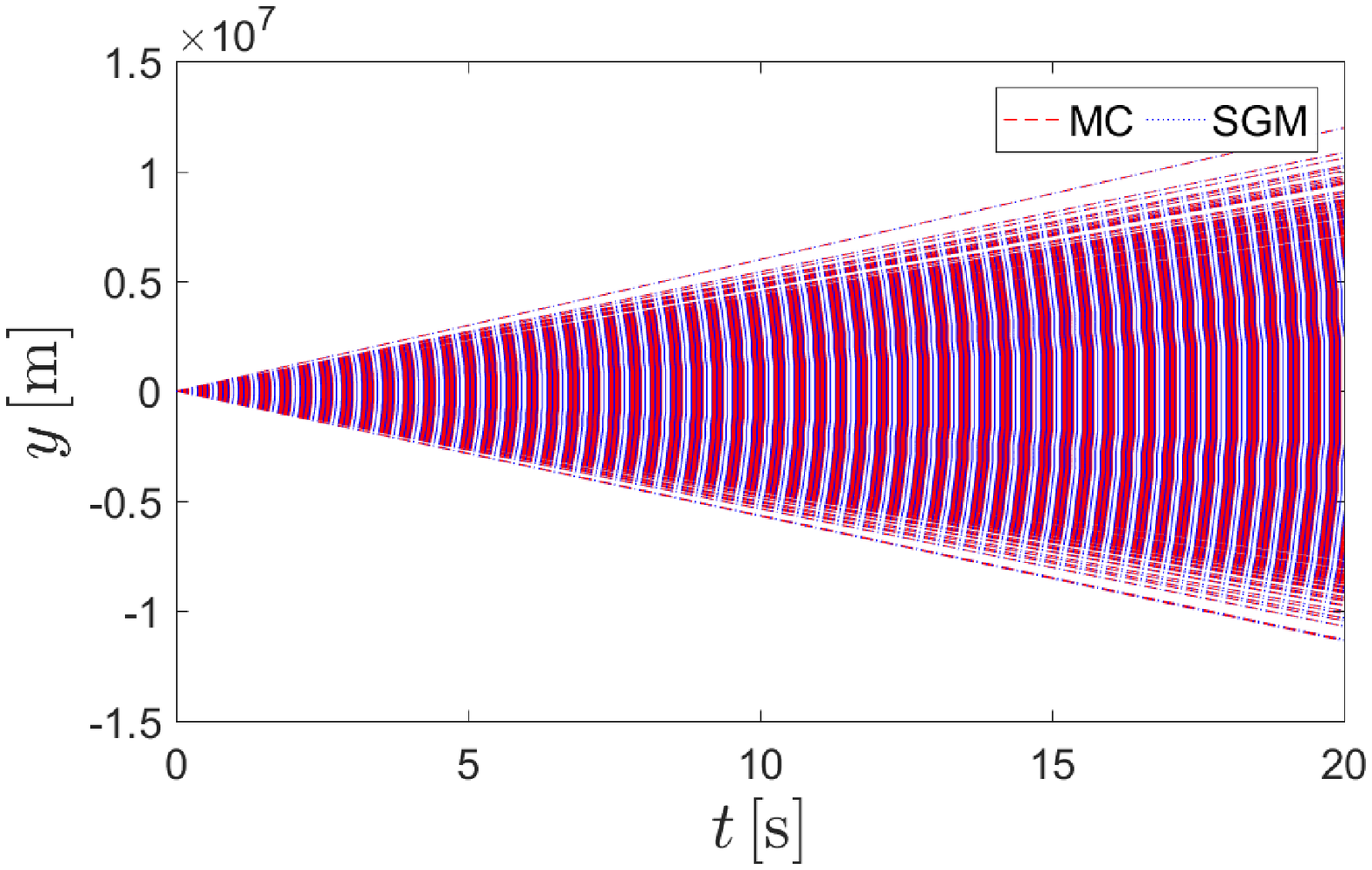}}
	\subfigure[Solution for $v_x(t)$.] {\includegraphics[width=0.45\textwidth]{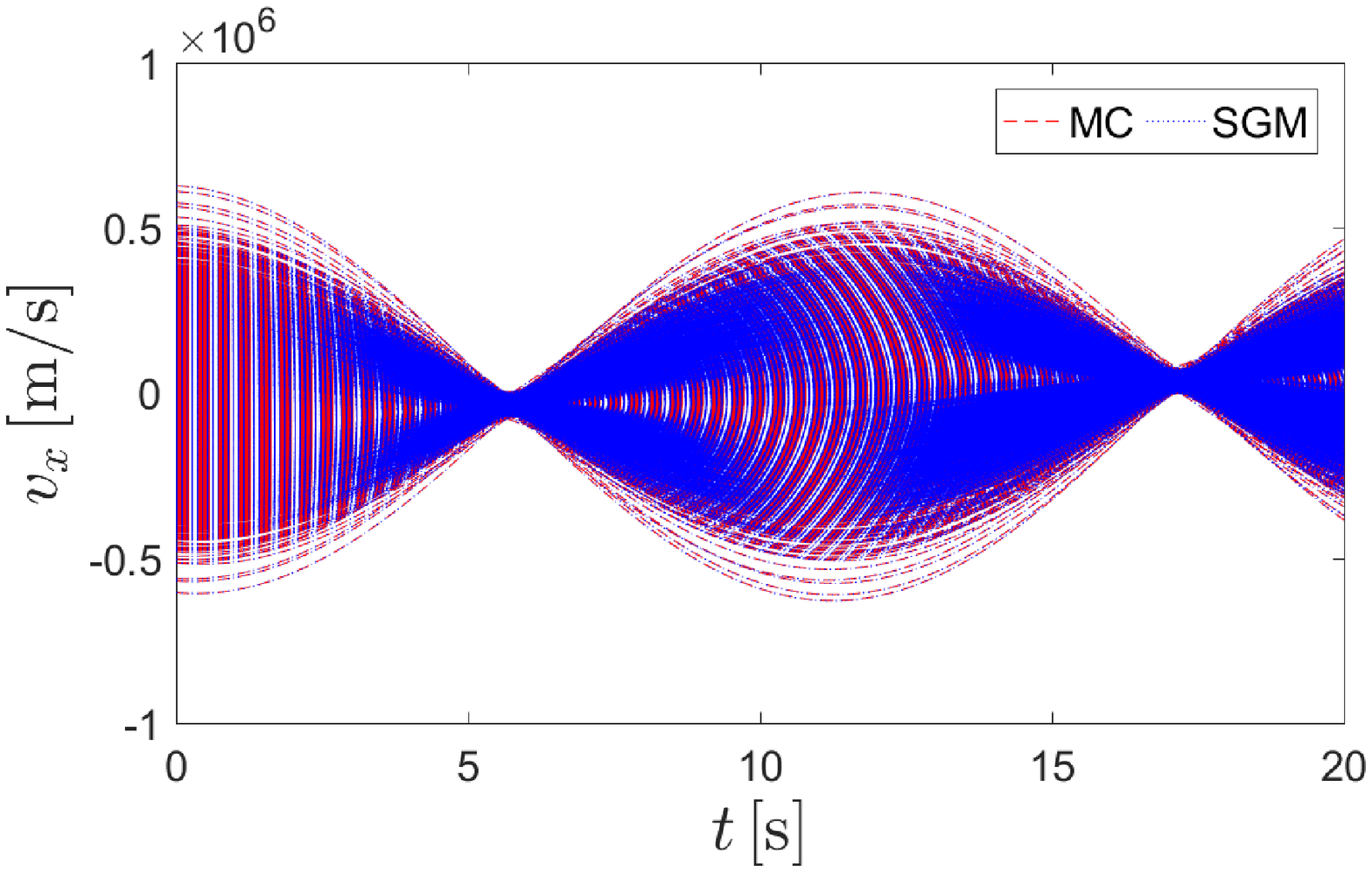}}
	\subfigure[Solution for $v_y(t)$.] {\includegraphics[width=0.45\textwidth]{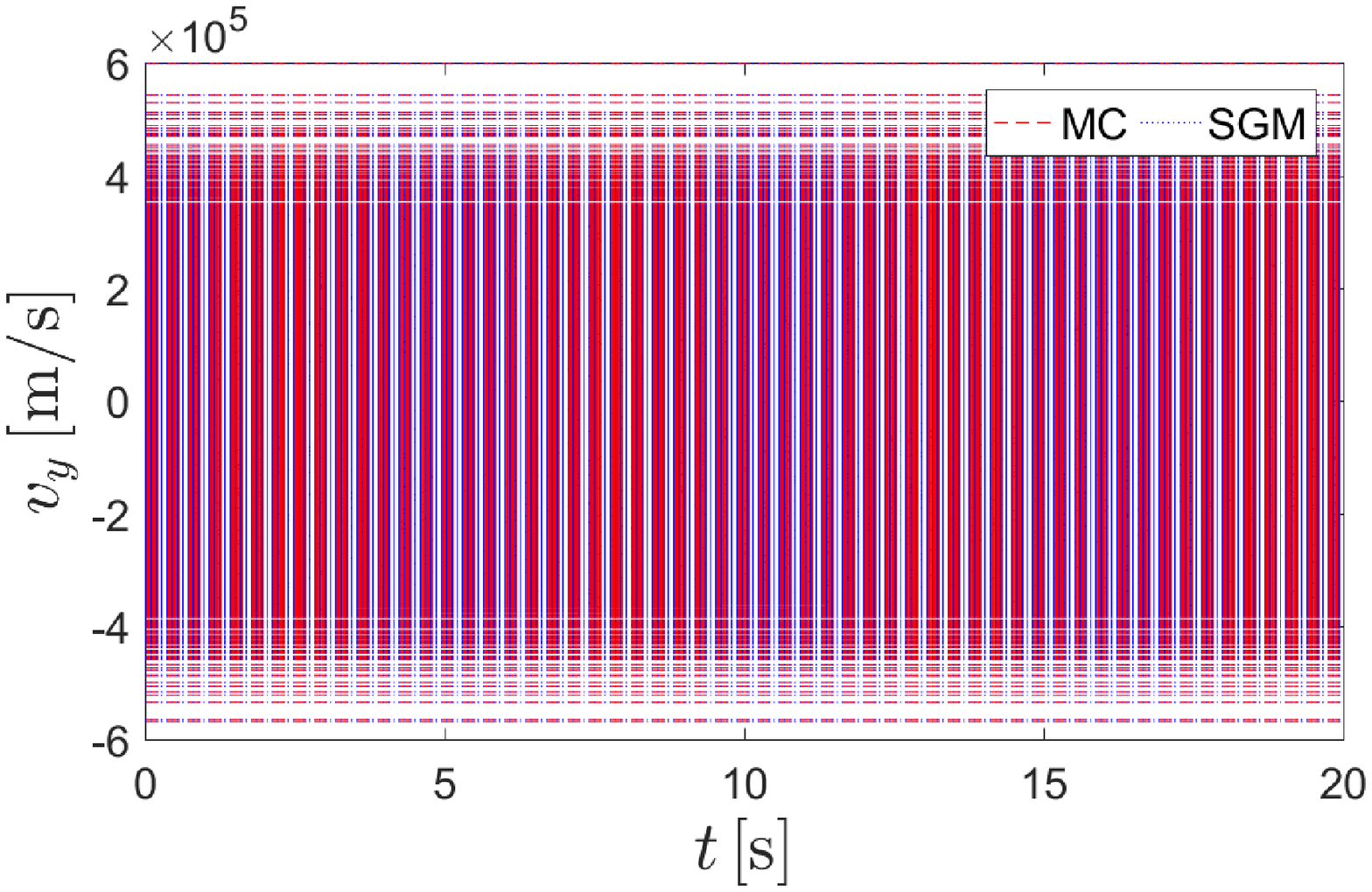}}
	\caption{\label{fig:tracking}Results of tracking of 10$^5$ particles. On the horizontal axis, the time in seconds over which the tracking simulation evolved is shown, and on the vertical axis, the resulting $x$, $y$, $v_x$ and $v_y$. The  dotted blue lines represent the solution computed using the Monte-Carlo simulations (MC), while the dashed red lines visualize the solutions of the stochastic Galerkin method (SGM). In all four cases, the positions of the lines are nearly indistinguishable, indicating the very good agreement between MC and SGM.}
\end{figure*}

When the dynamics includes electromagnetic fields, the stochastic expansion coefficients may evolve as a function of time and position. It is very common that the electromagnetic fields are functions of position, time, or frequency. This adds another level of complexity that the SGM must be capable to solve. As a consequence, the performance criterion in Eq.\,\ref{eq:err_time} must be modified to cope with position dependence as well,
\begin{subequations}
	\begin{align}
	\mathcal{\epsilon}_\mu(z)    & =	\bigg\lvert  \frac{\bar{\zeta}(z)-\bar{\hat{\zeta}}(z)}{\bar{\zeta}(z)}\bigg\rvert,\,\, \text{and} \\
	\mathcal{\epsilon}_\sigma(z) & =	\bigg\lvert  \frac{  \sigma[\zeta(z)]-\sigma[\hat{\zeta}(z)]}{\sigma[\zeta(z)]}\bigg\rvert.
	\end{align}\label{eq:err_space}
\end{subequations}
Here, $\zeta$ may refer either to the position, velocity or spin dependence, and $\hat{\zeta}$ constitutes the corresponding estimated value using the SGM, similar to Eq.\,(\ref{eq:err_time}). 

\section{Numerical results} \label{sec:res}
Two different simulation scenarios have been considered here. In the first scenario, described in Sec.\,\ref{sec:results:A}, uniform fields are used with realistic particle properties, while in the second scenario, described in Sec.\,\ref{sec:results:B}, a realistic beam passing through the fields of an RF Wien filter\,\cite{Slim2016116} that are computed numerically using the electromagnetic package CST MWS\footnote{Computer Simulation Technology, Microwave Studio, CST AG., Darmstadt, Germany, \url{http://www.cst.com}.}.   

\subsection{Uniform fields} \label{sec:results:A}
The generic simulation scenario serves as a \textit{proof-of-concept} demonstrator for the SGM. 10$^5$ particles are considered\footnote{Due to the limited computational resources to carry out the equivalent MC simulations, only 10$^5$ particles have been considered here. It should be noted that the SGM with the same resources can support the simulation of a much larger number of particles.} with normally-distributed initial positions and velocities traveling in a uniform electromagnetic field with $\vec{E}= (1,0,0)\,\si{V/m}$ and $\vec{H}= (0,1/173,0)\,\si{A/m}$\footnote{In vacuum, $\vec{B}=\mu_{0} \vec{H} = 4\cdot\pi \cdot 10^{-7} \vec{H}$.}, in a time interval from 0 to 20 seconds. 

The positions and transverse angles, $x$, $x'$, $y$ and $y'$ are generated using a 2$\sigma$ beam emittance of $\epsilon_{x,y} = \SI{1}{\mu m}$. The transverse velocities $v_x$ and $v_y$ are calculated by multiplying the transverse angles $x'$ and $y'$ by $v_z$. $v_z$ is also modeled as a Gaussian random variable with a mean value of $\beta \cdot c\,\,\si{m/s}$ and a standard deviation that corresponds to a variation of the beam momentum of $\Delta p/p = \num{e-4}$. The beam parameters used in the simulations are summarized in Table\,\ref{tab:beam}.  
\begin{table}[t]
\renewcommand{\arraystretch}{1.3}
	\begin{ruledtabular}
		\begin{tabular}{llr}
			Parameter                                           & Description         &              Value \\ \hline
			$N$                                                 & Number of particles &           \num{e5} \\
			$q$                                                 & Deuteron charge     &  \SI{1.602e-19}{C} \\
			$m$                                                 & Deuteron mass       & \SI{3,344e-27}{kg} \\
			$G$                                                 & Deuteron $G$-factor &       \num{-0.143} \\
			$c$                                                 & Speed of light      &  \SI{2.998e8}{m/s} \\
			$\beta = v/c$                                       & Lorentz $\beta$     &        \num{0.459} \\
			$\epsilon$                                          & Beam emittance      &    \SI{e-6}{\mu m} \\
			$\Delta p/p$                                        & Momentum variation  &          \num{e-4}
		\end{tabular}
	\end{ruledtabular}
	\caption{Typical parameters of a deuteron beam stored at COSY at a momentum of \SI{970}{MeV/c}, which are used in the particle and spin tracking simulations, described here.}
	\label{tab:beam}
\end{table}

Figure\,\ref{fig:tracking} shows the tracking results of the MC and the SGM-based simulation of $x$, $v_x$, $y$, and $v_y$. Due to the extended phase space distribution of the beam, the transverse velocities $v_x$ and $v_y$ do not vanish. In $x$-direction, this leads to a transverse Lorentz force and an oscillation of the particles around the beam direction, while in $y$-direction the particles are just drifting. 
It was verified by simulation that a beam of vanishing emittance $\epsilon$ \textit{and} momentum variation $\Delta p/p$, performs a perfect drift motion in both $x$- and $y$-directions.

Very good agreement between the MC method and the SGM-based simulations can clearly be  observed in Fig.\,\ref{fig:tracking}. In  particular, no difference between the oscillation periods and the magnitudes for the positions $x$ and the velocities $v_x$ are observed. As expected, both methods indicate a pure drift in radial direction, both in terms of position and velocity, as shown in Fig.\,\ref{fig:tracking} (d). 

\begin{figure}[t]
\centering
\includegraphics[width=\columnwidth]{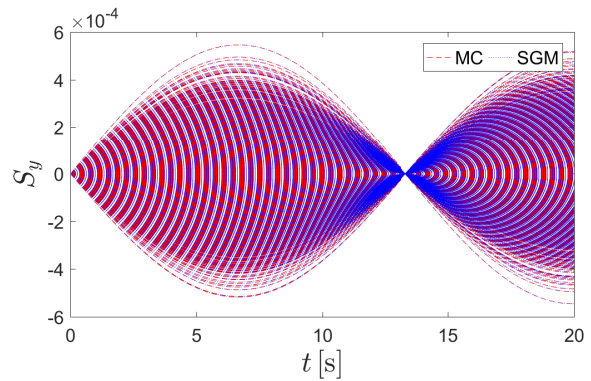}
\caption{Result of spin tracking simulations in uniform fields using the Monte-Carlo (MC) and the stochastic Galerkin method (SGM). The horizontal axis shows the time $t$ in seconds, and the vertical axis the vertical component $S_y(t)$ of the spin vector $\vec S(t)$.}
\label{fig:sy}
\end{figure}

The stochastic discretization of the T-BMT equation has been implemented as well, and the numerical results are shown in Fig.\,\ref{fig:sy}. The fields, despite the fact that they are uniform, can be expanded to include effects of undesired physical phenomena, such as, for instance, misalignments. The simulation scenario assumes a horizontally polarized deuteron beam with initial spin vectors in the horizontal (ring) plane, such that $\vec{S}=\left( 1, 0, 0 \right)$. While the system of ODEs is solved for $S_x$, $S_y$ and $S_z$, only the vertical spin component $S_y$ is displayed in Fig.\,\ref{fig:sy}. Of importance for the present investigation is the agreement between the SG and MC methods, and not a particular solution of the spin dynamics equations. This is clearly visible in Fig.\,\ref{fig:sy}, which shows that both oscillation magnitude and frequency of the MC and the SGM evolve synchronously.    
          
\begin{figure*}[t]
	\centering
	\subfigure[Error analysis for the mean value $\mathcal{\epsilon}_\mu(t)$ based on Eq.\,(\ref{eq:err_time}a).] {\includegraphics[width=0.4\textwidth]{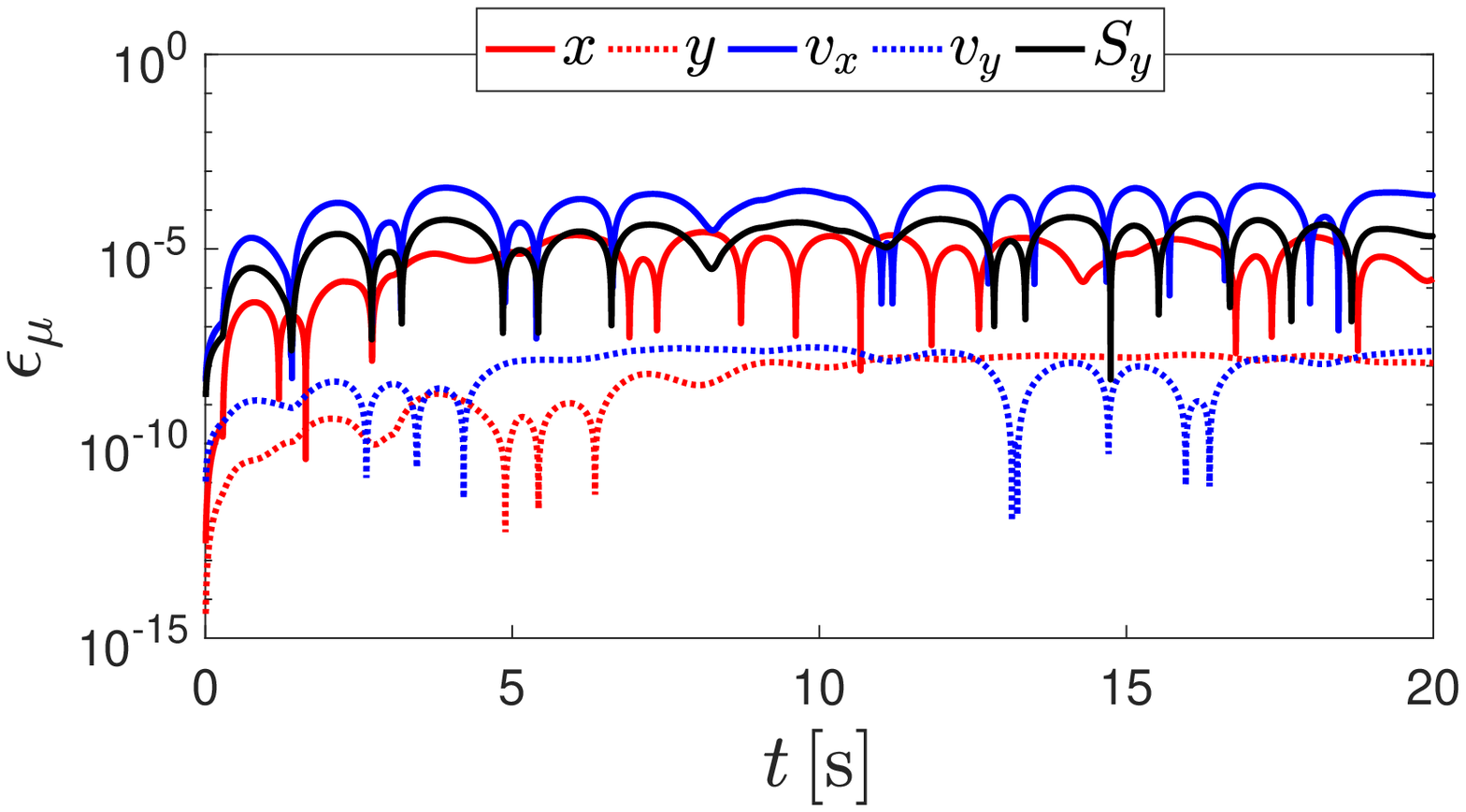}\label{fig:error_mu_gen}}
	\subfigure[Error analysis for the standard deviation $\mathcal{\epsilon}_\sigma(t)$ based on Eq.\,(\ref{eq:err_time}b).] {\includegraphics[width=0.4\textwidth]{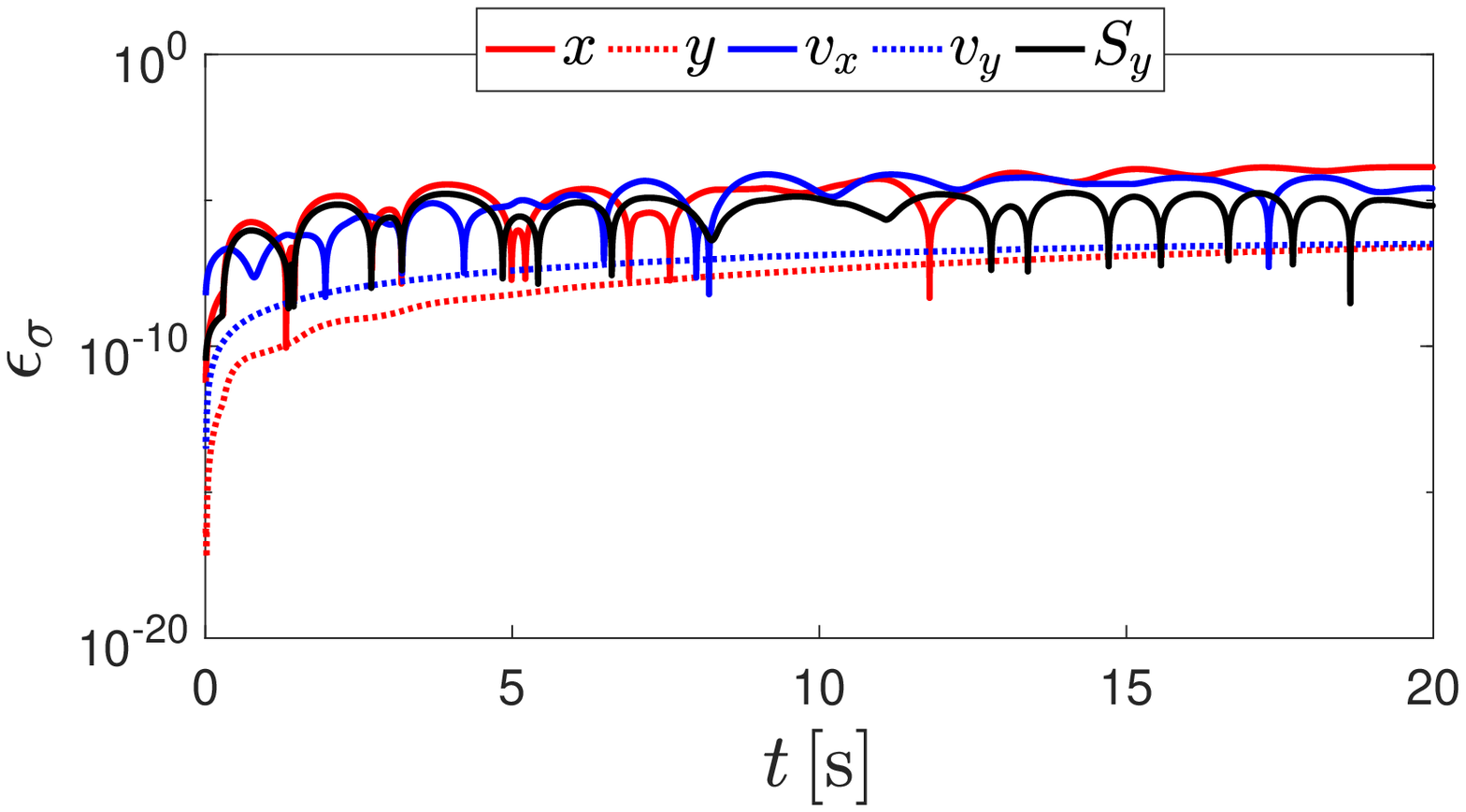}\label{fig:error_sigma_gen}}
	\caption{\label{fig:error_gen}Error analysis involving the mean value $\mathcal{\epsilon}_\mu(t)$ and the standard deviation $\mathcal{\epsilon}_\sigma(t)$ of the quantities $x$, $y$, $v_x$, $v_y$ and $S_y$  in the time interval from $t =0$ to $\SI{20}{s}$.}
\end{figure*}

By applying Eq.\,(\ref{eq:err_time}), the performance of the SGM is demonstrated in Figs.\,\ref{fig:error_mu_gen} and \,\ref{fig:error_sigma_gen}. As we are dealing with stochastic quantities, the mean and the standard deviations are used as performance indicators.  All estimated SGM solutions, $\hat{x}$, $\hat{v_x}$, $\hat{y}$, $\hat{v_y}$ and $\hat{S_y}$ do not deviate by more than \textbf{$10^{-5}$} from the MC solutions, $x$, $v_x$, ${y}$, ${v_y}$ and ${S_y}$, which was one of the goals of this study.
\begin{figure*}[t]
\centering
\subfigure[$z =\SI{1}{mm}$]{\includegraphics[width=0.24\textwidth]{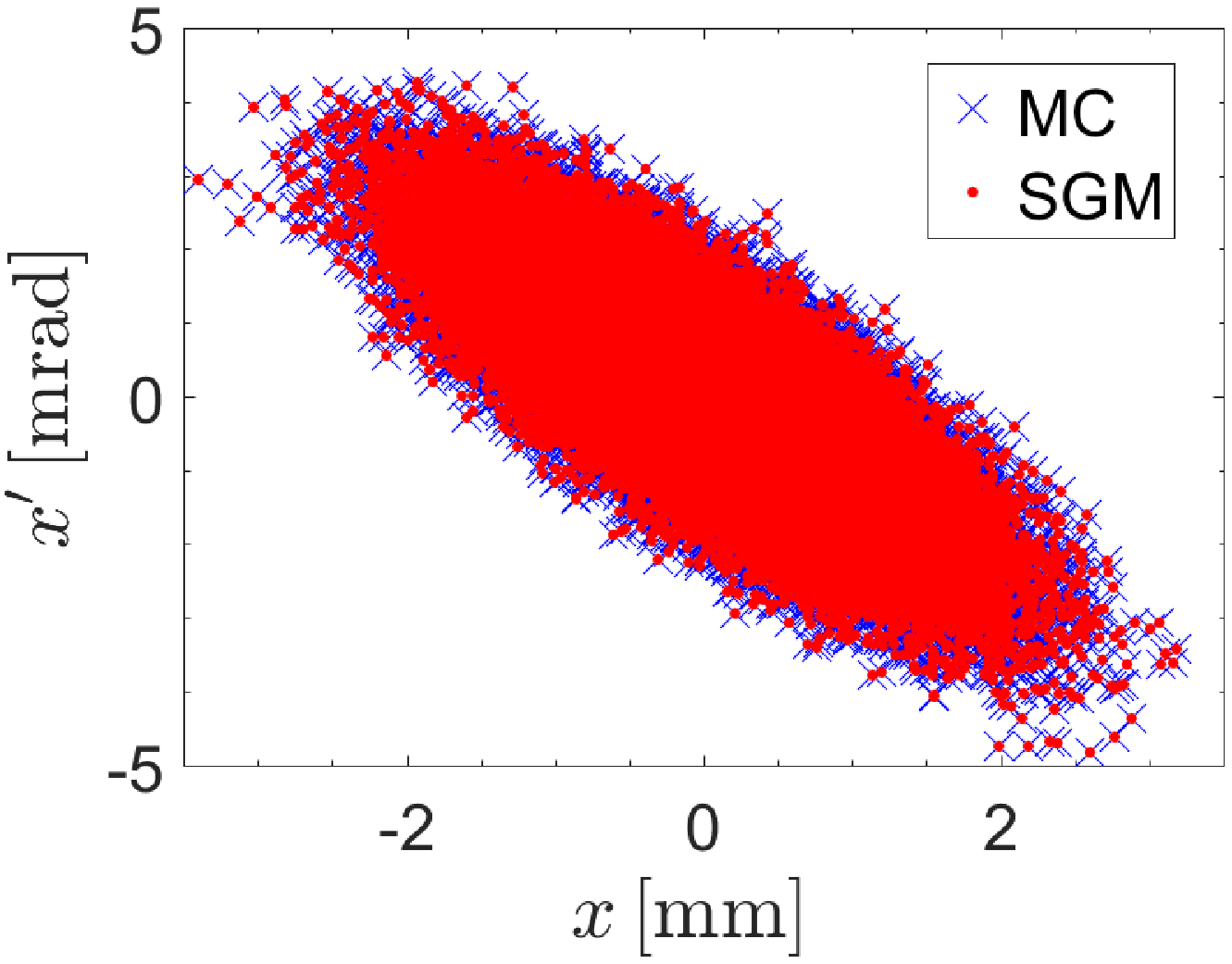}}
\subfigure[$z =\SI{210}{mm}$]{\includegraphics[width=0.24\textwidth]{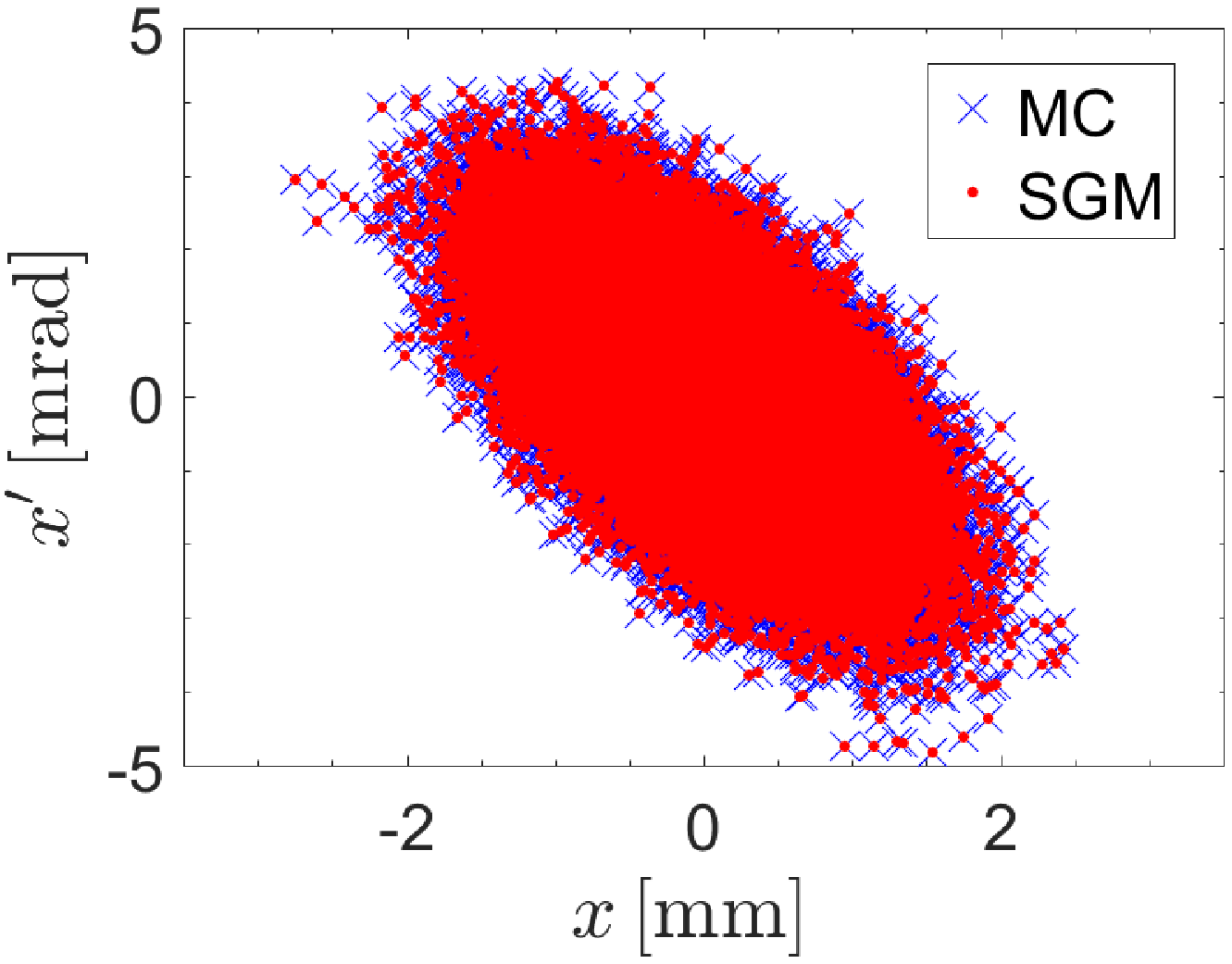}}
\subfigure[$z =\SI{610}{mm}$]{\includegraphics[width=0.24\textwidth]{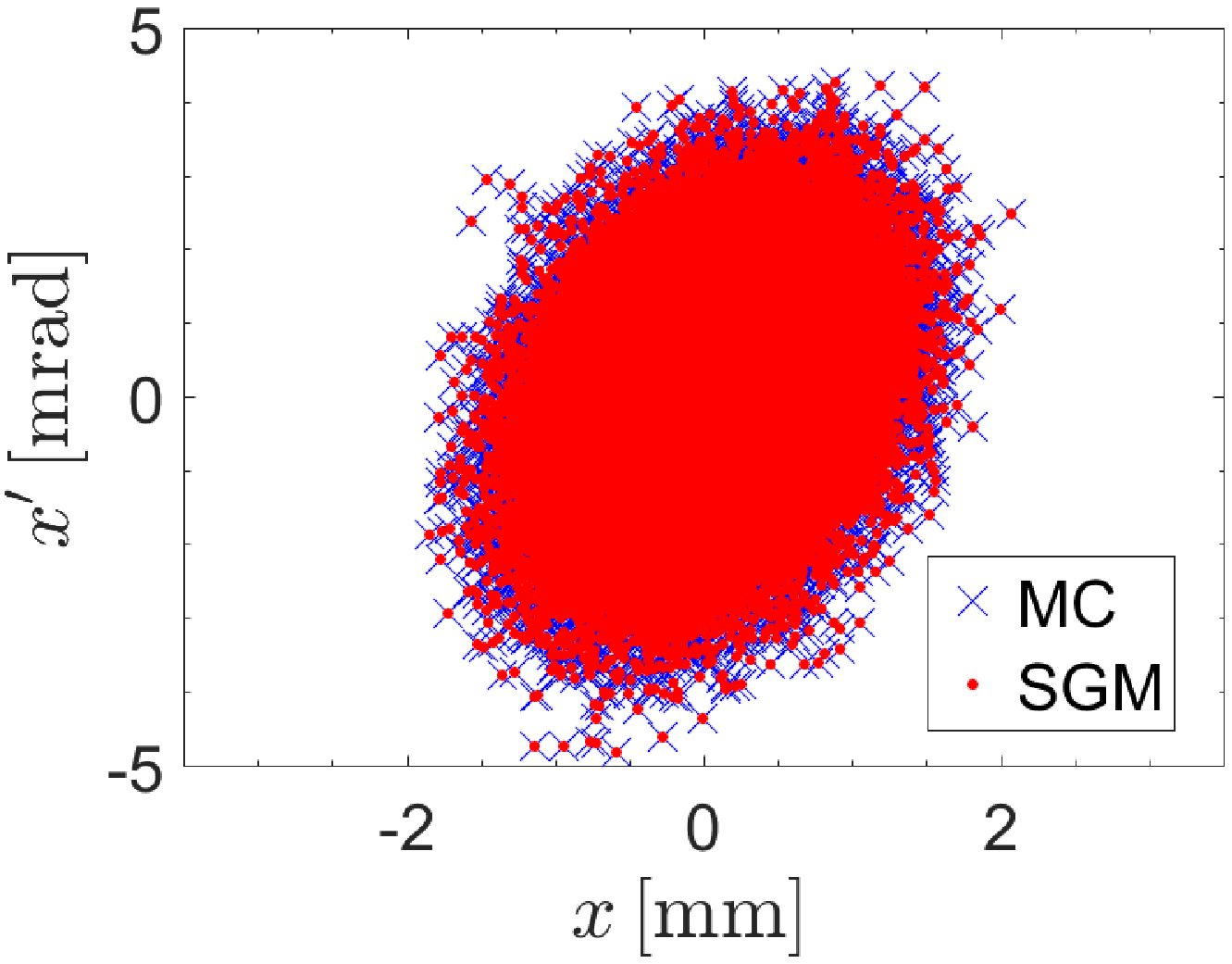}}
\subfigure[$z =\SI{810}{mm}$]{\includegraphics[width=0.24\textwidth]{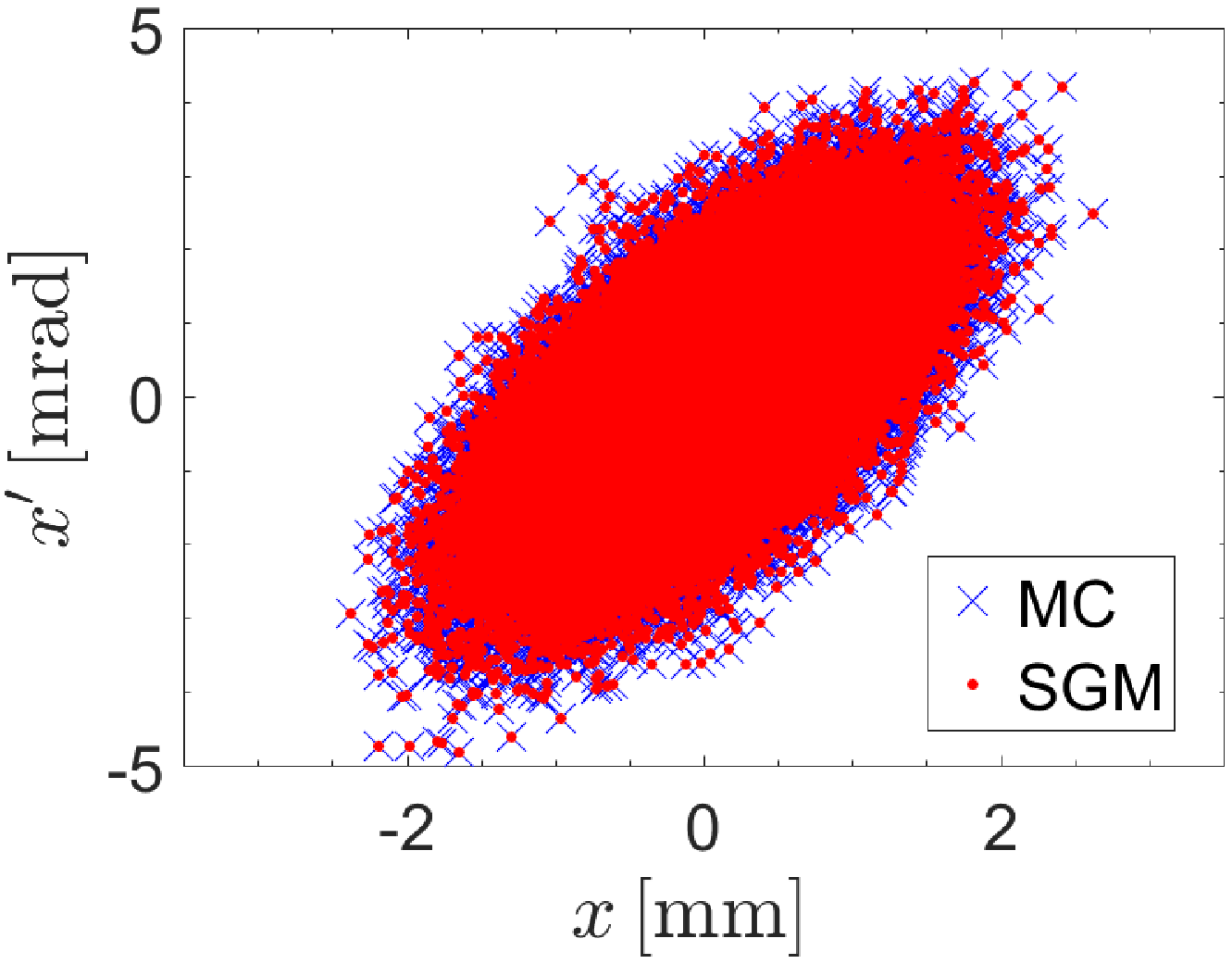}}
\caption{\label{fig:pse}Comparison between the phase-space ellipses between the Monte-Carlo (MC) tracking results (blue crosses) and the stochastic Galerkin method (SGM, red dots) at $ z = 1$, $210$, $610$ and $\SI{810}{mm}$ along the beam direction in the waveguide RF Wien filter\,\cite{Slim2016116}. The red dots are very dense so that the blue crosses cannot be seen anymore.}
\end{figure*}

\subsection{Tracking with realistic RF Wien filter fields} \label{sec:results:B}

In this section, we present simulation results using the MC and the SGM, where both methods were applied to deuteron beam (see Table\,\ref{tab:beam}) traveling along the $z$-axis of an RF Wien filter, whose electromagnetic fields were calculated using a full-wave simulation based on CST$^1$. A detailed description of the calculations and the field distributions can be found in\,\cite{Slim2016116}.

The waveguide RF Wien filter constitutes a novel device that is presently installed at COSY-J\"ulich aiming at a first direct measurement of the deuteron EDM in the framework of the JEDI collaboration\,\cite{Rathmann:2013rqa}. The device is characterized by high-quality electromagnetic fields that, when properly matched, provide a vanishing Lorentz force. This is achieved by adjusting the field quotient $Z_q$\footnote{The field quotient is defined as the ratio of total electric field to total magnetic field.}, which makes the device transparent to passage of particles (see Eq.\,(3) of\,\cite{Slim2016116}).

At the location of COSY, where the device is installed, the beam size is adjustable by modifying the $\beta$-function to values between about $\beta = 0.4$ and $\SI{4}{m}$\,\cite{PhysRevSTAB.18.020101}. In this paper, we consider only the case when $\beta = 0.4$, and a typical initial phase-space ellipse for a well-cooled beam at the entrance of the RF Wien filter at $z = \SI{1}{mm}$ are shown in panel (a) of Fig.\,\ref{fig:pse} (for other beam properties, see also Table\,\ref{tab:beam}).  In panels (b) to (d), the phase-space ellipses are displayed at the positions $z=210$, $610$ and $\SI{810}{mm}$. In all cases, SGM and MC results are in very good agreement.  It should furthermore be noted that for the purpose of the present paper, single-pass tracking calculations, \textit{i.e.}, without the ring, fully suffice to make the point.

\begin{figure}[htb]
\centering
\includegraphics[width=\columnwidth]{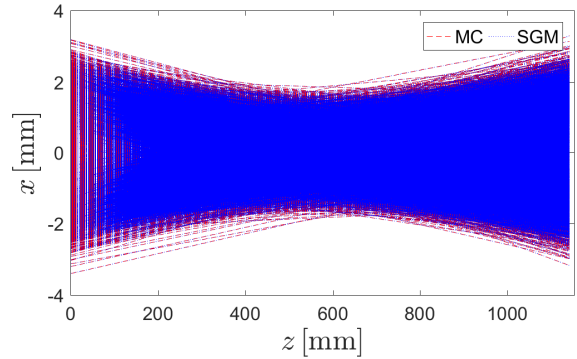}
\caption{Trajectory simulation of the $xz$ plane in the drift region along the RF Wien filter using the Monte-Carlo (MC) and the stochastic Galerkin method (SGM).}
\label{fig:x_sol_rfwf}
\end{figure}

The simulation results from the MC and SGM of the trajectories in the $xz$ plane of the drift region inside the RF Wien filter are shown in Fig.\,\ref{fig:x_sol_rfwf}. The MC results are represented by the dashed red lines, and the SGM simulations by the dotted red lines. Evidently, also here, the SGM results perfectly coincide with their MC counterparts.

\begin{figure}[htb]
	\centering
	\includegraphics[width=0.9\columnwidth]{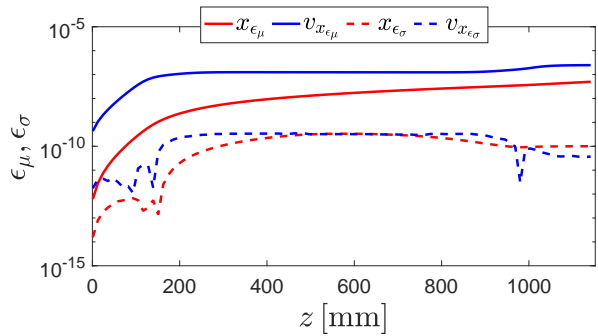}
	\caption{Error analysis involving the mean value $\mathcal{\epsilon}_\mu(z)$ and the standard deviation $\mathcal{\epsilon}_\sigma(z)$ involving the quantities $x$ and $v_x$  along the beam direction in the waveguide RF Wien filter.}
	\label{fig:error_rfwf}
\end{figure}

Using Eqs.\,(\ref{eq:err_space}a and b), the deviation of the simulation results of the SGM and the corresponding MC are quantified in Fig.\,\ref{fig:error_rfwf}. The horizontal axis represents the beam axis in the RF Wien filter. The solid lines denote the errors calculated by considering the mean values $\epsilon_{\mu}(z)$ with respect to the horizontal position $x$ and the velocity $v_x$ in $x$-direction, while the dashed lines correspond to the errors related to the standard deviation $\epsilon_{\sigma}(z)$ for the same quantities. The relative errors in the mean value sense of $x_{{\epsilon}_{\mu}}$ and $v_{x_{{\epsilon}_{\mu}}}$ do not exceed $10^{-7}$, while the error in the mean-squared sense of $x_{{\epsilon}_{\mu}}$ and $v_{x_{{\epsilon}_{\mu}}}$ do not exceed $10^{-10}$. The smallness of the calculated errors is a good indicator for the excellent performance of the SGM.  
\subsection{Comparison of simulation times}
One important performance criterion besides the error analysis is the comparison of the required simulation times. In the following, the generic scenario, described in Sec.\,\ref{sec:results:A}, is discussed, with particles properties as listed in Table \ref{tab:beam}.
 \begin{figure}[t]
 	\centering
 	\includegraphics[width=0.9\columnwidth]{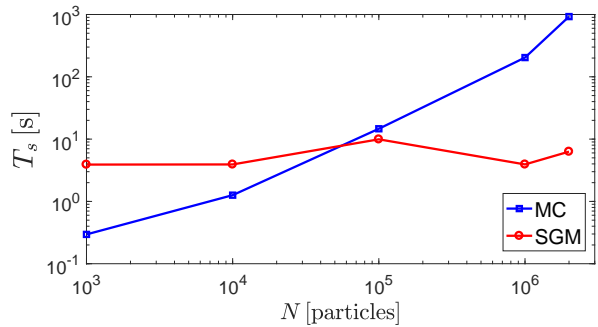}
 	\caption{Comparison of the simulation time required for the parallelized MC and the stochastic Galerkin method (SGM). For particle numbers below about \num{e5}, the methods are comparable, while for larger particle numbers, the SGM with a constant expansion order of $p=4$ performs much better than the MC. }\label{fig:run_time}
 \end{figure}
 
The MC simulations require to evaluate the system of beam and spin ODEs [Eqs.\,(\ref{eq:eom}, \ref{eq:eom_cart})]. The Matlab simulation environment provides a powerful vectorization option that has been used here to parallelize the code execution. As shown in Fig.\,\ref{fig:run_time}, for less than $N \approx \SI{e5}{particles}$, MC and SGM are about equally fast, but when $N$ increases further, the required time for the MC increases exponentially\footnote{The time required for the simulating of \num{2e6} particles on the available machine$^6$ took about $\num{925.42} \pm \SI{0.82}{s}$. Turning off the vectorization option, tracking of up to $10^9$ particles becomes possible, but the simulation time becomes prohibitively large.}, while the time required for the SGM stays about constant. 

Regardless of the number of particles, the system of ODEs involved in the SGM is evaluated exactly $P$ times\footnote{Even this number can be further reduced using a sparse version of PCE\,\cite{Slim201752}.}, corresponding to the number of basis functions. This is the main reason why the SGM is so much faster than the MC, without reduction in accuracy, as evidenced in Figs.\,\ref{fig:error_mu_gen}, \ref{fig:error_sigma_gen}, and \ref{fig:error_rfwf}. When the expansion order $p$ is kept constant, the number of basis functions remains constant as well, and hence also the simulation time $T_s = \SI{4.14}{} \pm \SI{0.39}{s}$, as shown in Fig.\,\ref{fig:run_time}.
\section{Conclusion} \label{sec:conc}
This paper reports on the application of the stochastic Galerkin method (SGM) to beam and spin tracking simulations. The method has been shown to work well with uniform fields. It has been applied to a realistic scenario, involving the electromagnetic fields of the waveguide RF Wien filter as well. The results indicate very good agreement with the Monte-Carlo simulations, that were carried out concurrently. 

The error calculations that we carried out show that the performance of the SGM is statistically equivalent to the  Monte-Carlo (MC) method, but with much lower computational demand. While the computational effort of MC-based simulations increases \textit{exponentially} as function of particle number, the computational effort involved in the SGM is constant, \textit{independent} of the number of particles tracked. The SGM transforms the original system of beam and spin ODEs into an augmented system of chaos coefficients, which are determined and then used to reconstruct the response for an arbitrary number of particles. The SGM is therefore capable to track large particle numbers in short time without compromising on the accuracy, and in this way provides a very efficient alternative to MC-based methods.  

A potential future application of the SGM might be  spin tracking calculations for storage rings, which are necessary in particular for precision experiments, such as the search for electric dipole moments. In such cases, an ultimate precision is required in the presence of uncertainties of the optical elements that constitute the machine. The SGM allows one to conveniently take into account systematic errors from different sources and to build a hierarchy of error sources. In view of the computational effort required for the MC-based error evaluation, the SGM may thus become an indispensable tool for future precision experiments.

\section*{Acknowledgment}
This work has been performed in the framework of the JEDI collaboration and is supported by an ERC Advanced-Grant of the European Union (proposal number 694340). We would kindly like to thank Alexander Nass, Helmut Soltner, Martin Gaisser, and J\"org Pretz, Andreas Lehrach and Jan Hetzel for their useful comments and careful reading of the manuscript.
\bibliographystyle{apsrev4-1}
\bibliography{dBase}
\end{document}